\documentclass[12pt]{article}
\usepackage[margin=0.9in]{geometry}
\usepackage{amsmath}
\usepackage{amssymb}
\usepackage{mathptmx}
\usepackage{graphicx,psfrag,epsf}
\usepackage{enumerate}
\usepackage{natbib}
\usepackage{url} 
\usepackage{multirow}
\usepackage{bm}
\usepackage{color}
\usepackage{floatrow}
\usepackage{units}
\usepackage{url}
\usepackage{verbatim}
\usepackage{slashbox}
\usepackage{epstopdf}
\usepackage{booktabs,caption,fixltx2e}
\usepackage[flushleft]{threeparttable}
\usepackage{rotating}
\usepackage{multirow}
\usepackage{physics}
\usepackage{placeins} 

\usepackage{fancyvrb}

\usepackage[linesnumbered,ruled]{algorithm2e}
\usepackage{nameref}

\newtheorem{theorem}{Theorem}

\newcommand{\DD}{\Delta\!\!\!\!\Delta}
\newcommand{\mbf}[1]{\mathbf{#1}}
\newcommand{\ind}{\stackrel{\mathrm{ind}}{\sim}}
\newcommand{\appsim}{\stackrel{\mathrm{a}}{\sim}}
\newcommand{\ippop}{\stackrel{P_{\theta_{0}}}{\rightarrow}}
\newcommand{\ipnu}{\stackrel{P_{\theta_{0}},P_{\nu}}{\rightarrow}}

\newcommand\smallO{
  \mathchoice
    {{\scriptstyle\mathcal{O}}}
    {{\scriptstyle\mathcal{O}}}
    {{\scriptscriptstyle\mathcal{O}}}
    {\scalebox{.7}{$\scriptscriptstyle\mathcal{O}$}}
  }
\newcommand{\Var}{\mathrm{Var}}

\newcommand{\blind}{1}

\begin{document}

\def\spacingset#1{\renewcommand{\baselinestretch}%
{#1}\small\normalsize} \spacingset{1}

\if1\blind
{
  \title{\bf Bayesian Uncertainty Estimation Under Complex Sampling}
  \author{Matthew R. Williams\hspace{.2cm}\\
    National Center for Science and Engineering Statistics\\
    National Science Foundation \\
    matthew.dunn.williams@gmail.com \\
    and \\
    Terrance D. Savitsky \\
    U.S. Bureau of Labor Statistics \\
    Savitsky.Terrance@bls.gov}
  \maketitle
} \fi

\if0\blind
{
  \bigskip
  \bigskip
  \bigskip
  \begin{center}
    {\LARGE\bf Bayesian Uncertainty Estimation Under Complex Sampling}
\end{center}
  \medskip
} \fi

\bigskip

\begin{abstract}
Social and economic studies are often implemented as complex survey designs. For example, multistage, unequal probability sampling designs utilized by federal statistical agencies are typically constructed to maximize the efficiency of the target domain level estimator (e.g., indexed by geographic area) within cost constraints for survey administration. Such designs may induce dependence between the sampled units; for example, with employment of a sampling step that selects geographically-indexed clusters of units.  A sampling-weighted pseudo-posterior distribution may be used to estimate the population model on the observed sample.  The dependence induced between co-clustered units inflates the scale of the resulting pseudo-posterior covariance matrix that has been shown to induce under coverage of the credibility sets.  By bridging results across Bayesian model mispecification and survey sampling, we demonstrate that the scale and shape of the asymptotic distributions are different between each of the pseudo-MLE, the pseudo-posterior and the MLE under simple random sampling. Through insights from survey sampling variance estimation and recent advances in computational methods,  we devise a correction applied as a simple and fast post-processing step to MCMC draws of the pseudo-posterior distribution. This adjustment projects the pseudo-posterior covariance matrix such that the nominal coverage is approximately achieved. We make an application to the National Survey on Drug Use and Health as a motivating example and we demonstrate the efficacy of our scale and shape projection procedure on synthetic data on several common archetypes of survey designs.
\end{abstract}

\noindent%
\vfill

\newpage
\spacingset{1.45} 

\section{Introduction}\label{sec:intro}
We focus on the task of the data analyst to estimate a Bayesian model, $P_{\theta_{0}}$, that they suppose generates values for a random variable, $Y$, for units of a population, $U = (1,\ldots,N)$, from an observed sample, $S=(1,\ldots,n\leq N)$, drawn from that population under a complex sampling design governed by distribution, $P_{\nu}$. Common designs include one-stage samples of businesses and multistage samples of households. These two major classes of sampling designs cover the majority of survey designs and represent a standard data collection method for social and economic data. While our methods and simulations will apply to both of these and to other types, we will focus on multistage designs for our motivating example.

In this section we introduce the concepts and current state of the literature for implementing Bayesian models for survey data. We also provide an introduction to variance estimation for the survey-weighted maximum likelihood estimator and to model mis-specification in Bayesian models. As we will demonstrate, combining results across these different areas provides an opportunity to innovate and improve on current methods. In particular, we can formally demonstrate a deficiency in the coverage of Bayesian credible intervals using the popular survey-weighted pseudo-posterior as well as provide an efficient and effective correction, achieving nominal coverage asymptotically and in practice for moderate sample sizes. With this adjustment, the pseudo-posterior now possesses many desirable theoretical properties. Unlike other methods in the literature, the pseudo-posterior can be implemented across a very wide variety of survey designs and analyst-specified population models. Even though the adjustment is asymptotic, the resulting posterior still retains its small sample characteristics (i.e. is not forced to be asymptotically normal).

\subsection{Informative Sampling Designs}
Multistage sampling designs are created to achieve efficient (low variance) estimation of a desired simple quantile, mean, or total estimator for a collection of domains within constraints on cost to administer the survey.   A first stage of the sampling design often collects contiguous geographic areas from the population into clusters, where a subset of the clusters are randomly selected into the sample for this first stage.  The contiguity of areas within each cluster is defined for convenience and cost to collect the sample, but it induces a dependence among units nested in areas within each cluster.  Dependencies among sampled units may be additionally promulgated through the drawing of a fixed-sized sample, without replacement, in any stage of the sampling design; for example, by constructing a systematic sampling step with a fixed interval using a random starting point, as is common when sampling households on the same street.

The sampling design distribution is induced by specifying marginal inclusion probabilities at each stage.  Survey agencies, such as Federal statistical agencies, publish marginal inclusion probabilities for last-stage sampled units, $\pi_{i} = Pr\left\{\delta_{i} = 1\right\}\in(0,1]$, for (observed) units, $i = 1,\ldots,n$, sampled in the last stage of the sampling design, where $n$ denotes the number of units in the observed sample and $\delta_{i}\in\{0,1\}$ specifies a unit inclusion indicator.

Efficiency of the population estimator, $g(Y)$, is enhanced through designing the inclusion probabilities, $(\pi_{i})_{i=1,\ldots,N}$, to be correlated with $(y_{i})_{i=1,\dots,N}$, where $N$ denotes the size of population, $U$; an example is the use of a proportion-to-size sampling design in the Current Employment Statistics (CES) survey of business establishments, administered by the U.S. Bureau of Labor Statistics, for the purpose of measuring total employment by geographic area and industry.  Higher unit inclusion probabilities are assigned to larger employers because they drive the variance of the resulting total employment estimator. Sampling designs that induce this correlation are termed ``informative" and the balance of information in the sample is different from that in the population.

Analyzing the data ``as is'' assuming a simple random sampling mechanism will then lead to bias. The most common approach in the survey literature is to use a survey-weighted likelihood. This is a plug-in estimator that formulates a sampling-weighted pseudo-likelihood density by exponentiating each (last stage) unit-indexed likelihood contribution by a sampling weight constructed to be inversely proportional to the unit marginal inclusion probability, $w_{i} \propto  1/\pi_{i}$, where $\pi_{i} = P\left(\delta_{i} = 1\right)$, for units, $i = 1,\ldots,n$, where $n$ denotes the number of units in the observed sample.  An approximate, weight-exponentiated pseudo-likelihood for the population, $\mathop{\prod}_{i = 1}^{n}p\left(y_{i}\vert \theta \right)^{w_{i}}$, is constructed from the $n$ units observed in the sample. The survey weights mitigate the estimation bias, but the uncertainty distribution (covariance structure) also needs to be estimated and adjusted.

\subsection{Variance Estimation for Survey-Weighted Maximum Likelihood}
When establishing the consistency of the survey-weighted pseudo-mle $\hat{\theta}_{ML}$,  it is almost ubiquitous in the literature \citep{Isaki82, pfeffermann98, chambers2003analysis} to assume that the pairwise inclusion probabilities $\pi_{ii'} = Pr\left\{\delta_{i} = 1, \delta_{i'} = 1\right\}\in(0,1]$ factor or are independent asymptotically ($\pi_{ii'} = \pi_{i} \pi_{i'}$). However, the typical assumption for variance estimation (of the same models) is to assume an arbitrary amount of within cluster dependence both in the sampling design and the population generating model \citep{heeringa2010applied, Rao92}. Recent work \citep{2018dep} provides the theoretical conditions and motivating examples to extend our understanding  of consistency to these real-world situations for which there are already well-established procedures for estimating variance.

\cite{Wang2017} suggest estimating variance based on the Horvitz-Thompson estimator for weighted sums. Theoretically this is appealing, but in practice it requires knowledge of second order inclusions probabilities $\pi_{ii'}$. These joint probabilities are rarely calculated and are disseminated even less frequently. Instead, the de-facto approach for variance estimation is based on the approximate sampling independence of the primary sampling units \citep{heeringa2010applied}. Variance estimation can be in the form of Taylor linearization or replication based methods. We provide a high-level overview of the two below, but a variety of implementations are available \citep{binder1996, Rao92}. Let $y_{ij}$, $X_{ij}$, and $w_{ij}$ be the observed data for individual $i$ in cluster $j$ of the sample. Assume the parameter $\theta$ is a vector of dimension $d$ with population model value $\theta_0$.

\begin{enumerate}
\item Taylor Linearization
	\begin{enumerate}
	\item Approximate an estimate $\hat{\theta}$, or a `residual' $(\hat{\theta} - \theta_{0})$, as a weighted sum:  $\hat{\theta} \approx \sum_{i,j} w_{ij} z_{ij}(\theta)$ where $z_{ij}$ is a function evaluated at the current values of $y_{ij}$, $X_{ij}$, and $\hat{\theta}$.
	\item Compute the weighted components for each primary cluster (e.g., primary sampling units (PSUs)):
	$\hat{\theta}_j = \sum_{i} w_{ij} z_{ij} (\theta)$.
	\item Compute the variance between clusters:
		$\widehat{Var(\hat{\theta})} = \frac{1}{J-d}\sum_{j =1}^{J}(\hat{\theta} -\hat{\theta}_j )(\hat{\theta} -\hat{\theta}_j )^{T}$
	\item For stratified designs, compute $\hat{\theta_s}$ and $\widehat{Var(\hat{\theta_s})}$ within strata and sum $\widehat{Var(\hat{\theta})} = \sum_{s} \widehat{Var(\hat{\theta_s})}$.
	\end{enumerate}
\item Replication
	\begin{enumerate}
	\item Through randomization (bootstrap), leave-one-out (jackknife), or orthogonal contrasts (balanced repeated replicates), create a set of $K$ replicate weights $(w_{i})_k$ for all $i \in S$ and for every $k = 1,\ldots, K$.
	\item Each set of weights has a modified value (usually $0$) for a subset of clusters, and typically has a weight adjustment to the other clusters to compensate:
	$\sum_{i\in S} (w_{i})_k = \sum_{i\in S} w_{i}$ for every~$k$.
	\item Estimate $\hat{\theta}_k$ for each replicate $k\in 1,\ldots, K$.
	\item Compute the variance between replicates:
		$\widehat{Var(\hat{\theta})} = \frac{1}{K-d}\sum_{k =1}^{K}(\hat{\theta} -\hat{\theta}_k )(\hat{\theta} -\hat{\theta}_k )^{T}$.
	\item For stratified designs, generate replicates such that each strata is represented in every replicate.
	\end{enumerate}
\end{enumerate}
There are two notable challenges associated with these methods:
\begin{itemize}
	\item For Taylor linearization, the value of $\hat{\theta}$ is typically only computed \emph{once} and then used in a plug in value for $z_i (\theta)$. Whereas for the replication methods, the estimate $\hat{\theta}_k$ must be computed $K$ times. This may lead to a sizable differences in \emph{computational effort} for models of moderate complexity and a moderate number of replicates $K$.
	\item For the replication methods, no additional derivatives as needed. In contrast, the Taylor linearization method often requires the calculation of a gradient based on the estimating equation for $\hat{\theta}$ to derive the analytical form of the first order approximation $z_i (\theta)$. This poses significant \emph{analytical challenges} for all but the simplest models.
\end{itemize}

In this work we present a third option which can be implemented as a hybrid of the two approaches. This approach can then be applied to any sampling design, as long as the analyst has either the replication weights or the cluster and stratification information. The implementation (Section \ref{sec:adjustment}) is made possible by recent advances in algorithmic differentiation \citep{Margossian18}, which allows us to specify the model as a log density but only treat the gradient in the abstract \emph{without} specifying it analytically.

\subsection{Bayesian Models for Survey Data}

\subsubsection{Generalized Etimating Equations and Method of Moments}
\cite{yin2009} propose using the normal distribution with parameters from the solution of generalized method of moments to generate MCMC posterior draws. An analogous approach with survey-weighted estimating equations was proposed and implemented by \cite{shah2000} to estimate a logistic mixed effects model for survey data, where the fixed effects were sampled from a normal distribution with mean estimates from a survey-weighted likelihood maximization and variance estimates from a Taylor series linearization approach. \cite{Wang2017} formalizes the use of the generalized method of moments approach for survey data and provides a Bernstein-von Mises result, demonstrating that the resulting posterior intervals achieve correct frequentist coverage asymptotically. These methods assume asymptotic normality when making posterior draws. While large sample performance is generally good, small sample properties and performance are uncertain.  By contrast, our approach addresses fully Bayesian estimation performed by the data analyst under a model of their choosing.  The data analyst will extract numerical draws from the joint distribution for their model parameters, from which a variety of statistics of interest may be computed (e.g., quantiles, probabilities for events).   Our method performs a rescaling of draws from the marginal or joint posterior distributions obtained from the model and MCMC algorithm specified by the data analyst to produce credibility sets with asymptotically correct frequentist coverage.   The approaches of \cite{shah2000,Wang2017}, by contrast, require the availability of an argmax of the posterior (MAP) point estimate obtained from an estimating equation, which may not be reliably computable for a complicated hierarchical Bayesian model specification.   These approaches also circumvent the estimation of the fully Bayesian model specified by the data analyst, so they may not address the small sample borrowing of strength provided by a fully Bayesian specification as does our method.

\subsubsection{Likelihood Approaches}

\citet{2015arXiv150707050S} proposed a plug-in estimator that formulates a sampling-weighted pseudo-posterior density that is analogous to the weighted pseudo-likelihood. However, the sampling weights, $(w_{i})$, are normalized, to $(\tilde{w}_{i})$, to control the amount of estimated posterior uncertainty. \citet{2015arXiv150707050S} default to normalizing, $\sum_{i=1}^{n}\tilde{w}_{i} = n$.
\citet{2017arXiv171000019N} demonstrate that the pseudo-posterior estimator constructed from weights normalized to $n$ generally produce credibility intervals that fail to contract on frequentist confidence sets by under covering because they don't account for dependencies among units induced by the joint distribution $(P_{\theta_{0}},P_{\nu})$.
\cite{2017arXiv171000019N} develop an alternative approach to the pseudo-posterior distribution that multiplicatively adjusts the likelihood to accomplish asymptotically unbiased estimation of the population model on the observed informative sample.  This extends the formulation of the observed likelihood by \cite{pfeffermann98} to a fully Bayesian implementation by specifying a conditional population model, $p(\pi_{i}\mid y_{i},\kappa)$, for the inclusion probabilities, $(\pi_{i})_{i\in U}$.
\citet{Pfef:DaS:DoN:mult:2006} also extend \citet{pfeffermann98} to a partially Bayesian estimation, but they treat $(\pi_{i})_{i\in U}$ as \emph{fixed}. So the approach of \citet{Pfef:DaS:DoN:mult:2006} may be viewed to be \emph{not} fully Bayesian, because it does not specify a joint or conditional model for $p(\pi_{i}|y_{i}, \kappa)$.

\cite{2017arXiv171000019N} show that credible intervals estimated from their adjusted, fully Bayes posterior achieves correct coverage in the case of a simple, single stage proportion to size sampling design.   Their likelihood adjustment, however, requires a different MCMC sampler than that developed for the population model and the adjusted likelihood includes an integration that must be numerically computed in each MCMC draw.  So the fully Bayesian estimator lacks the broad ease-of-implementation of the pseudo-posterior approach.

\citet{raowu2010} also address the under coverage of pseudo-posterior in the specific case of formulating $P_{g(Y)}$ as an empirical likelihood for the purpose of estimating a total or mean, $\widehat{g(Y)}$.  They replace $n$ as the normalizer for $(\tilde{w}_{i})$ with $n^{\ast} = n / \mbox{DEFF}_{\widehat{g(Y)}}$, where $\mbox{DEFF}_{\widehat{g(Y)}} = \Var_{P_{\nu}}(\widehat{g(Y)}) / \Var_{\mbox{SRS}}(\widehat{g(Y)})$  denotes the design effect, defined as the variance induced under sampling design distribution, $P_{\nu}$, divided by that under simple random sampling (SRS).  Their approach improves the coverage properties for estimation of simple statistics, rather than some $\theta$ of interest to the data analyst for a general $P_{\theta}$.  In addition, the simultaneous modeling of multiple outcomes or parameters would require multiple DEFF's to be used, which is not possible if DEFF is only incorporated via scaling the sample size~$n$.

In the sequel we demonstrate that a post-processing adjustment step applied to the pseudo-posterior MCMC samples corrects the under coverage demonstrated by \cite{2017arXiv171000019N}.  The fully Bayes approach will tend to be produce more efficient credible sets, however, under the requirement to specify a conditional population model for the inclusion probabilities that is \emph{assumed} to be correctly specified. In practice, sample designs are often algorithmically defined, becoming quite complex. The fully Bayes approach has not been applied to multistage cluster designs. The impact of clustering on the effective sample size may still be a challenge. In this work, we demonstrate that the survey-weighted pseudo-posterior can be adjusted to give correct inference even under complex survey designs which include within-cluster dependence.

\subsection{Mis-specification for Bayesian Models}
When a Bayesian model has a mis-specified likelihood, the resulting posterior distribution contracts on an alternative distribution which is the minimum Kullback-Leibler distance from the true generating distribution \citep{kleijn2012}. However \cite{kleijn2012} demonstrate that Bayesian credible sets from the posterior are not valid confidence sets. In other words, under mis-specification, the MLE and the posterior distribution for the mis-specified model have different limiting distributions. This suggests that an adjustment is needed to achieve valid confidence sets asymptotically.
\citet{2009arXiv0911.5357R} motivate a similar sandwich form of an adjustment of the asymptotic covariance of the pseudo-posterior distribution under specification of a composite weight-exponentiated pseudo-likelihood, where their pseudo-likelihood is employed to approximate a likelihood that is not able to be specified. \citet{2009arXiv0911.5357R} redesign the MCMC sampler to accomplish the adjustment,  such that their approach requires the development of a specialized MCMC sampler, distinct from the sampler developed for the population, $P_{\theta_{0}}$.

Our survey sampling formulation assumes existence of a population model, $P_{\theta_{0}}$, which, though unknown, has a tractable form that allows consistency of our estimator, $P_{\theta}$. Even though consistency is achieved, the survey-weighted pseudo-posterior is still mis-specified because the exponentially weighted likelihood is a noisy approximation to the true likelihood of the joint distribution $(P_{\theta_{0}},P_{\nu})$.
The sampling-weighted pseudo-posterior arises out of a random sampling mechanism to approximate the information in the population using a partially observed sample taken from that population. In contrast, \citet{2009arXiv0911.5357R} don't compute expectations with respect to the joint distribution, $(P_{\theta_{0}},P_{\nu})$, to develop their adjustment since they do not contemplate a random sampling process governed by $P_{\nu}$. We provide theoretical results for the form of the asymptotic sampling-weighted pseudo-MLE covariance matrix under the joint distribution for population generation and the taking of a sample.  We also provide a clean post-estimation adjustment that allows for minimal changing of a user's intended MCMC implementation.

\subsection{Adjusting the Distribution of the Pseudo-Posterior} 
The current work constructs a simple post-processing step that adjusts the scale and shape of sampling-weighted, pseudo-posterior parameter credibility sets that we show in the sequel achieves approximately correct coverage under a broad class of generally-used sampling designs.  Our procedure applies an adjustment step to the posterior draws to achieve an asymptotic sandwich form for the pseudo-posterior covariance that is the same as that for the sampling-weighted pseudo-MLE. We accomplish the adjustment by computing the variance of the score function and the expectation of the square of its gradient under the joint distribution, $(P_{\theta_{0}},P_{\nu})$. The variance of the gradient is estimated via a hybrid approach combining the principles of Taylor linearization and replication. The implementation (Section \ref{sec:adjustment}) is through the use of algorithmic differentiation \citep{Margossian18} and randomized replication sampling \citep{Preston09}.

The adjustment step is applied, numerically, by resampling the observed data, $y_{1},\ldots,y_{n}$, under an empirical distribution approximation for $(P_{\theta_{0}},P_{\nu})$.  The re-sampling step is implemented by simply drawing blocks of units from the existing sample at those stages where dependence is induced within the blocks.  All units nested within each re-sampled block are included in each re-sample; for example, if the multistage design includes a clustering step, we use the known cluster memberships of the last stage units and just re-sample the clusters.
The population generating distribution, $P_{\theta_{0}}$, is estimated, once, on our original sample and the adjustment is evaluated using the best available estimate for $\theta$, the posterior mean.  Our adjustment is, therefore, computationally fast and achieves nearly correct coverage for $\theta$.  The pseudo-posterior MCMC sampler, used for estimation of $\theta$, requires only a simple edit to the population posterior sampler (to insert sampling weights) because the same posterior geometry is employed.  Our adjustment procedure requires \emph{no} change to the MCMC sampler for the pseudo-posterior, which preserves its ease of use.

\section{Motivating Multistage Cluster Design:  The National Survey on Drug Use and Health}
Our motivating survey design is the National Survey on Drug Use and Health (NSDUH), sponsored by the Substance Abuse and Mental Health Services Administration (SAMHSA). NSDUH is the primary source for statistical information on illicit drug use, alcohol use, substance use disorders (SUDs), mental health issues, and their co-occurrence for the civilian, non institutionalized population of the United States.
The NSDUH employs a multistage state-based design
\citep{MRB:Sampling:2014},
with the earlier stages defined by geography within each state in order to select households (and group quarters) nested within these geographically-defined primary sampling units (PSUs).
\citet{2018dep} provides conditions for asymptotic consistency for the pseudo-posterior for designs like the NSDUH, which are characterized by:
\begin{itemize}
\item Cluster sampling, such as selecting only one unit per cluster, or selecting multiple individuals from a dwelling unit.
\item Population information such as socio-economic indicators used to sort sampling units along gradients.
\end{itemize}
Both features are common, in practice, and create sampling dependencies that do not attenuate even if the population grows. For simplicity of exposition we examine the relationship between two measures, current (past month) smoking of cigarettes and past year major depressive episode for adults through a two-parameter logistic regression model. Both cigarette smoking and depression may be clustered geographically and within households. For example, rates for each tend to vary by age, urban/rural status, education, and other demographics which typically cluster geographically and within household \citep{DT:2014, MHDT:2014}.

\section{Asymptotic Covariance of the Pseudo-Posterior Distribution}
\label{asymptotics}
\subsection{The Pseudo-Posterior Framework}
We suppose random variables of the population are generated, $\mathbf{X}_{\nu} = \left(\mbf{X}_{1},\ldots,\mbf{X}_{N_{\nu}}\right) \ind P_{\theta_{0}}$ where $\theta_{0} \in \mathbb{R}^{d}$ and we perform inference on $\theta \in \Theta$ of the population model from the sample of size, $n_{\nu}$.
A sampling design imposes a \emph{known} distribution on a vector of random inclusion indicators, $\bm{\delta}_{\nu} = \left(\delta_{\nu 1},\ldots,\delta_{\nu N_{\nu}}\right)$, on units composing a population, $U_{\nu}$.  The sampling distribution takes an \emph{observed} random sample, $S_{\nu} \subseteq U_{\nu}$, of size $n_{\nu} \leq N_{\nu}$ from $U_{\nu}$. Our conditions for the main results are based on marginal unit inclusion probabilities, $\pi_{\nu i} = \mbox{Pr}\{\delta_{\nu i} = 1\}$ for all $i \in U_{\nu}$ and the second order pairwise probabilities, $\pi_{\nu ij} = \mbox{Pr}\{\delta_{\nu i} = 1 \cap \delta_{\nu j} = 1\}$ for $i,j \in U_{\nu}$, which are obtained from the joint distribution over $\left(\delta_{\nu 1},\ldots,\delta_{\nu N_{\nu}}\right)$. We denote the sampling distribution by $P_{\nu}$, which governs the taking of samples from the population. $P_{\nu}$ is implicitly conditionally defined given realizations from $P_{\theta_{0}}$. In other words, the joint distribution for $\bm{\delta}_{\nu}$ can depend on some population information from $\mathbf{X}_{\nu}$.

We denote the \emph{observed} sample of size $n_{\nu}$ as $\{\mbf{X}_{\nu},\bm{\delta}_{\nu} \} = \left(\{\mbf{X}_{\nu 1},\delta_{\nu 1}\},\ldots,\{\mbf{X}_{\nu N_{\nu}},\delta_{\nu N_{\nu}}\}\right)$ , following \citet{2015arXiv150707050S,SJOS:SJOS12312,2018dep}, where $\delta_{\nu i} = 0$ indicates unit $i$ is \emph{not} included in the sample removes the associated $\mbf{X}_{\nu i}$.  It is a notational convention that emphasizes the dependence of generated samples on both $P_{\nu}$ (which governs $\delta_{\nu i}$) and $P_{\theta_{0}}$ (which governs the generation of population values, $\mbf{X}_{\nu i}$).   Since the $\bm{\delta}_{\nu}$ are random with respect to $P_{\nu}$, $\{\mbf{X}_{\nu},\bm{\delta}_{\nu} \}$ is jointly random with respect to $(P_{\nu}, P_{\theta_{0}})$.

The inclusion probabilities are formulated to depend on the finite population data values, $\mathbf{X}_{\nu}$, so that we employ the pseudo-posterior estimator to approximate the population likelihood from the observed sample with,
\begin{equation} \label{pseudolike}
p^{\pi}\left(\mbf{X}_{\nu i},\delta_{\nu i}\right) := p\left(\mbf{X}_{\nu i}\right)^{\delta_{\nu i}/\pi_{\nu i}},~i \in U_{\nu},
\end{equation}
which weights each density contribution, $p(\mbf{X}_{\nu i})$, by the inverse of its marginal inclusion probability \citep{2015arXiv150707050S}. When $\delta_{\nu i} = 0$, the pseudo-posterior likelihood contribution for unit $i$ under conditional independence (given $\theta$) is removed.  This approximation for the population likelihood produces the associated pseudo-posterior density,
\begin{equation}\label{inform_post}
p^{\pi}\left(\theta  \mid \mbf{X}_{\nu},\bm{\delta}_{\nu}  \right) = \frac{\mathop{\prod}_{i\in U_{\nu}}p^{\pi}_{\theta}\left(\mbf{X}_{\nu i},\delta_{\nu i}\right)\pi(\theta)}{\mathop{\int}_{\Theta}\mathop{\prod}_{i\in U_{\nu}}p^{\pi}_{\theta}\left(\mbf{X}_{\nu i},\delta_{\nu i}\right)\pi(\theta)d\theta},
\end{equation}
where $\mbf{X}_{\nu},\bm{\delta}_{\nu} = \left(\mbf{X}_{\nu 1},\delta_{\nu 1},\ldots,\mbf{X}_{\nu N_{\nu}},\delta_{\nu N_{\nu}}\right)$ denotes the observed sample of size, $n_{\nu}$. The pseudo-posterior mass placed on subset $B \subseteq \Theta$ becomes
\begin{equation}\label{postmass}
\Pi^{\pi}\left(B\big\vert\mbf{X}_{\nu},\bm{\delta}_{\nu}\right) = \mathop{\int}_{\theta\in B}p^{\pi}\left(\mbf{X}_{\nu},\bm{\delta}_{\nu} \mid  \theta  \right)\pi(\theta)d\theta
\end{equation}

In typical applications \citep{SJOS:SJOS12312}, sampling weights are normalized to satisfy $ \mathop{\sum}_{i\in S_{\nu}} \pi^{-1}_{\nu i} = n_{\nu}$, which regulates the scale of uncertainty in the estimated pseudo-posterior distribution.  In practice, dependencies induced by informative, multistage sampling designs produce a smaller effective sample size than $n_{\nu}$, such that the typical procedure under-estimates posterior uncertainty.  In addition, the shape (geometry) of the pseudo-posterior distribution is impacted by the dependence induced in each stage of the sampling design such that the asymptotic covariance matrix will not be the same as that for the MLE obtained under simple random sampling.  We proceed to derive the form of the limiting covariance matrix for the pseudo-MLE under informative sampling, which we define as the MLE of Equation~\ref{pseudolike}.  We demonstrate that the covariance matrix of the pseudo-MLE is different from that for the MLE under simple random sampling, but that the latter is a special case of the former.  We next demonstrate that the limiting covariance matrix of the pseudo-posterior distribution differs from the pseudo-MLE under informative sampling (due to the failure of Bartlett's second identity) such that resulting credibility intervals would not be expected to contract on valid frequentist confidence intervals, absent adjustment.

The difference between the limiting covariance matrix for the pseudo-posterior distribution, on the one hand, from that for the MLE under simple random sampling, on the other hand, may only be partly driven by informativeness of the sampling design. The dependencies induced under employment of a multistage sampling design, such as the within cluster dependence of units, will also impact the scale of the limiting covariance matrix of the pseudo-posterior distribution, even absent sampling informativeness.  In other words, even where sampling inclusion probabilities, $(\pi_{\nu i})$, are not required to provide unbiased estimation of $\theta\in\Theta$ (i.e. the design is `ignorable'), the resulting limiting covariance matrix of the posterior distribution under multistage sampling would be different from that for the MLE under simple random sampling.

Our main result is achieved in the limit as $\nu\uparrow\infty$, under the countable set of successively larger-sized populations, $\{U_{\nu}\}_{\nu \in \mathbb{Z}^{+}}$.  The asymptotics under our construction is controlled by $\nu\in\mathbb{N}$ to map to the process where we fix a $\nu$, construct an associated finite population of size, $N_{\nu}$, generate random variables $\mbf{X}_{\nu 1},\ldots, \mbf{X}_{\nu N_{\nu}} \ind P_{\theta_{0}}$, construct unit marginal sample inclusion probabilities, $(\pi_{\nu 1},\ldots,\pi_{\nu N})$ (and other design features such as cluster and strata identifiers) under $P_{\nu}$ and then draw a sample, $\{1,\ldots,n_{\nu}\}$ from that population.  The process is repeated for each increment of $\nu$.  We define the associated stochastic rates of convergences notations, $A_{\nu} = \smallO_{P}\left(B_{\nu}\right)$ to denote that $A_{\nu} = Y_{\nu}B_{\nu}$ where $\displaystyle Y_{\nu} \mathop{\rightarrow}^{P} 0$ and $A_{\nu} = \mathcal{O}_{P}\left(B_{\nu}\right)$ denotes $A_{\nu} = Y_{\nu}B_{\nu}$ where $Y_{\nu} = \mathcal{O}_{P}\left(1\right)$.  For deterministic sequences, $A_{\nu}$ and $B_{\nu}$, the notations reduce to the usual $\smallO$ and $\mathcal{O}$.

\subsection{Review of Survey-Weighted Empirical Functionals}\label{prelim}
We use the empirical distribution approximation for the joint distribution over population generation and the draw of an informative sample that produces our observed data.  Our empirical distribution construction follows \citet{breslow:2007} and incorporates inverse inclusion probability weights, $\{1/\pi_{\nu i}\}_{i=1,\ldots,N_{\nu}}$, to account for the informative sampling design,
\begin{equation}
\mathbb{P}^{\pi}_{N_{\nu}} = \frac{1}{N_{\nu}}\mathop{\sum}_{i=1}^{N_{\nu}}\frac{\delta_{\nu i}}{\pi_{\nu i}}\delta\left(\mbf{X}_{\nu i}\right),
\end{equation}
where $\delta\left(\mbf{X}_{\nu i}\right)$ denotes the Dirac delta function, with probability mass $1$ on $\mbf{X}_{\nu i}$ and we recall that $N_{\nu} = \vert U_{\nu} \vert$ denotes the size of the finite population. This construction contrasts with the usual empirical distribution, $\mathbb{P}_{N_{\nu}} = \frac{1}{N_{v}}\mathop{\sum}_{i=1}^{N_{\nu}}\delta\left(\mbf{X}_{\nu i}\right)$.

We will construct asymptotic distributions for the sequence of centered and scaled random quantities,
\begin{equation}
h_{N_{\nu}} = \sqrt{N_{\nu}}\left(\theta - \theta_{0}\right),
\end{equation}
for specific estimators.  Let $\hat{\theta}_{\pi,N_{\nu}} = \mathop{\arg\max}_{\theta}\mathop{\sum}_{i\in U_{\nu}} \frac{\delta_{\nu i}}{\pi_{\nu i}}\log \{p_{\theta}(X_{\nu i})\}$ denote the MLE of the pseudo-likelihood in Equation~\ref{pseudolike} (that we denote as the pseudo-MLE).  It is the MLE of the logarithm of the sample-weighted likelihood for an observed sample (where $\delta_{\nu i} = 1$ for those units included in the observed sample).  The pseudo-MLE defines the sequence,
\begin{equation}\label{eq:sampcenterscale}
\hat{h}^{\pi}_{N_{\nu}} = \sqrt{N_{\nu}}\left(\hat{\theta}_{\pi,N_{\nu}} - \theta_{0}\right),
\end{equation}
as contrasted with centered and scaled sequence for the MLE, $\hat{\theta}_{N_{\nu}}$, for the population (as if fully-observed),
\begin{equation}
\hat{h}_{N_{\nu}} = \sqrt{N_{\nu}}\left(\hat{\theta}_{N_{\nu}} - \theta_{0}\right).
\end{equation}
Define the log-likelihood, $\ell_{\theta} = \log p_{\theta} = \log p_{\theta_{0}+ h_{N_{\nu}}/\sqrt{N_{\nu}}} $ and the associated score function, $\displaystyle\dot{\ell}_{\theta} = \nabla_{\theta}\ell_{\theta}$.  Equation~\ref{eq:sampcenterscale} is scaled by $N_{\nu}$ because we later sum empirical expecations and variances over $N_{\nu}$ sampling-weighted units in the population, where $\bm{\delta}_{\nu}$ indexes all possible realizable samples \citep{breslow:2007}.

We follow the notational convention of \citet{Ghosal00convergencerates} and define the associated expectation functionals with respect to these empirical distributions by $\mathbb{P}^{\pi}_{N_{\nu}}f = \frac{1}{N_{\nu}}\mathop{\sum}_{i=1}^{N_{\nu}}\frac{\delta_{\nu i}}{\pi_{\nu i}}f\left(\mbf{X}_{\nu i}\right)$.  Similarly, $\mathbb{P}_{N_{\nu}}f = \frac{1}{N_{\nu}}\mathop{\sum}_{i=1}^{N_{\nu}}f\left(\mbf{X}_{\nu i}\right)$.  Lastly, we use the associated centered empirical processes, $\mathbb{G}^{\pi}_{N_{\nu}} = \sqrt{N_{\nu}}\left(\mathbb{P}^{\pi}_{N_{\nu}}-P_{0}\right)$ and $\mathbb{G}_{N_{\nu}} = \sqrt{N_{\nu}}\left(\mathbb{P}_{N_{\nu}}-P_{0}\right)$.

We construct two variance expressions, starting with Fisher's information:
\begin{equation}
H_{\theta_{0}} = -\frac{1}{N_{\nu}}\mathop{\sum}_{i\in U_{\nu}}\mathbb{E}_{P_{\theta_{0}}}\ddot{\ell}_{\theta_{0}}(\mbf{X}_{\nu i}),
\end{equation}
whose inverse provides the asymptotic covariance of the pseudo-posterior under our Bernstein Von-Mises result that follows. Next, we define:
\begin{equation}
J_{\theta_{0}} = \frac{1}{N_{\nu}}\mathop{\sum}_{i\in U_{\nu}}\mathbb{E}_{P_{\theta_{0}}}\dot{\ell}_{\theta_{0}}(\mbf{X}_{\nu i})\dot{\ell}_{\theta_{0}}(\mbf{X}_{\nu i})^{T},
\end{equation}
which is the middle term in the asymptotic variance of the MLE. Under the population model (and an SRS subsample), the likelihood is properly specified, so $J_{\theta_{0}} = H_{\theta_{0}}$.

Because our pseudo-posterior framework arises from a random sampling process governed by $P_{\nu}$,
\begin{align*}
 H^{\pi}_{\theta_{0}} &=-\mathbb{E}_{P_{\theta_{0}},P_{\nu}}\left[\mathbb{P}^{\pi}_{N_{\nu}}\ddot{\ell}_{\theta_{0}}\right] \\
 &= -\frac{1}{N_{\nu}}\mathop{\sum}_{i\in U_{\nu}}\mathbb{E}_{P_{\theta_{0}}}\left[\mathbb{E}_{P_{\nu}}\frac{\left[\delta_{\nu i}\vert \mathcal{A}_{\nu}\right]}{\pi_{\nu i}}\ddot{\ell}_{\theta_{0}}(\mbf{X}_{\nu i})\right]\\
&= -\frac{1}{N_{\nu}}\mathop{\sum}_{i\in U_{\nu}}\mathbb{E}_{P_{\theta_{0}}}\ddot{\ell}_{\theta_{0}}(\mbf{X}_{\nu i})\\
&= H_{\theta_{0}},
\end{align*}
where $\mathcal{A}_{\nu}$ denotes the sigma field of information in $U_{\nu}$.
We note that this equivalance between  $H^{\pi}_{\theta_{0}}$  and $H_{\theta_{0}}$ does not hold for the weighted composite likelihood of \citet{2009arXiv0911.5357R}, where the weights are arbitrary and arise from a deterministic process to approximate an intractable likelihood for the population, $U_{\nu}$.

Our main results in the following section are anchored in the observation that the survey-weighted $J^{\pi}_{\theta_{0}} =\mathbb{E}_{P_{\theta_{0}},P_{\nu}}\left[\mathbb{P}^{\pi}_{N_{\nu}}\ddot{\ell}_{\theta_{0}}\ddot{\ell}_{\theta_{0}}^{T}\right] \ne J_{\theta_{0}}$ due to the mis-specification from using a noisy approximation to the likelihood for $(P_{\theta_{0}},P_{\nu})$.

\subsection{Main Results}
The following conditions guarantee three results on the forms for asymptotic covariance matrices of the distributions for pseudo-MLE estimator and the pseudo-posterior.  The first theorem extends Theorem $5.23$ of \citet{vdv00} to derive the asymptotic expansion of the centered and scaled pseudo-MLE. The second theorem specifies the form of the associated sandwich covariance matrix for the (asymptotic expansion of the) pseudo-MLE.  The third theorem extends similar theorems in \citet{kleijn2012} and \citet{vdv00} that specify the covariance matrix of the asymptotic Gaussian form for the pseudo-posterior distribution.  We observe that the asymptotic covariance matrices are different for each of the MLE, the pseudo-MLE and the pseudo-posterior, which sets up our proposed scale and shape adjustment, introduced in the sequel.

\begin{description}
\item[(A1)\label{continuity}] (Continuity)
        For each $\theta\in\Theta\in\mathbb{R}^{d}$ (an open subset of Euclidean space), $\ell_{\theta_{0}}\left(\mbf{x}\right)$ be a measurable function (of $\mbf{x}$) and differentiable at $\theta_{0}$ for $P_{\theta_{0}}-$ almost every $\mbf{x}$ (with derivative, $\dot{\ell}_{\theta_{0}}\left(\mbf{x}\right)$), such that for every $\theta_{1}$ and $\theta_{2}$ in a neighborhood of $\theta_{0}$ with $\mathbb{E}_{\theta}\dot{\ell}_{\theta_{0}}\left(\mbf{x}\right)\dot{\ell}_{\theta_{0}}\left(\mbf{x}\right)^{T} < \infty$, we have a Lipschitz condition:
        \begin{equation*}
        \Big\vert\ell_{\theta_{1}}\left(\mbf{x}\right)  - \ell_{\theta_{2}}\left(\mbf{x}\right)\Big\vert \leq \dot{\ell}_{\theta_{0}}\left(\mbf{x}\right)\norm{\theta_{1}-\theta_{2}} \mbox{a.s.}~ P_{\theta_{0}}
        \end{equation*}
\item[(A2)\label{expansion}] (Local Quadratic Expansion)
        The Kullback-Liebler divergence with respect to $P_{\theta_{0}}$ has a second order Taylor expansion about $\theta_{0}$,
        \begin{equation*}
        \mathbb{E}_{P_{\theta_{0}}} \log\frac{p_{\theta}}{p_{\theta_{0}}} = \frac{1}{2}\left(\theta-\theta_{0}\right)^{T}H_{\theta_{0}}\left(\theta-\theta_{0}\right) + \smallO\left(\norm{\theta-\theta_{0}}^{2}\right),
        \end{equation*}
        where $H_{\theta_{0}}$ is a $d\times d$ positive definite matrix.
\item[(A3)\label{bartlettfirst}] (Bartlett's First Identity)
        \begin{equation*}
        \mathbb{E}_{P_{\theta_{0}}}\dot{\ell}_{\theta_{0}} = 0
        \end{equation*}
 \item[(A4)\label{consistency}] (Consistency of the MLE for the population)
        \begin{equation*}
        \mathbb{P}_{N_{\nu}}\ell_{\hat{\theta}_{N_{\nu}}} \geq \mathop{\sup}_{\theta}\mathbb{P}_{N_{\nu}}\ell_{\theta} - \smallO_{P_{\theta_{0}}}\left(N_{\nu}^{-1}\right)
        \end{equation*}
        and $\hat{\theta}_{N_{\nu}} \ippop \theta_{0}$
\item[(A5)\label{bounded}] (Non-zero Inclusion Probabilities)
        \begin{equation*}
        \displaystyle\mathop{\sup}_{\nu}\left[\frac{1}{\displaystyle\mathop{\min}_{i \in U_{\nu}}\vert\pi_{\nu i}\vert}\right] \leq \gamma, \text{  with $P_{\theta_{0}}-$probability $1$.}
        \end{equation*}
\item[(A6)\label{deprestrict}] (Growth of Dependence is Restricted)\\
For every $U_{\nu}$ there exists a binary partition $\{S_{\nu 1}, S_{\nu 2}\}$ of the set of all pairs $S_{\nu}= \{\{i,j\}: i\ne j \in U_{\nu}\}$ such that
        \begin{equation*}
         \displaystyle\mathop{\limsup}_{\nu\uparrow\infty} \left\vert S_{\nu 1} \right\vert \le \mathcal{O}\left(N_{\nu}\right),
        \end{equation*}
        and
 	 \begin{equation*}
        \displaystyle\mathop{\limsup}_{\nu\uparrow\infty} \mathop{\max}_{i,j \in S_{\nu 2}}\left\vert\frac{\pi_{\nu ij}}{\pi_{\nu i}\pi_{\nu j}} - 1\right\vert \le \mathcal{O}\left(N_{\nu}^{-1}\right), \text{  with $P_{\theta_{0}}-$probability $1$}
        \end{equation*}
\item[(A7)\label{fraction}] (Constant Sampling fraction)
        For some constant, $f \in(0,1)$, that we term the ``sampling fraction",
        \begin{equation*}
        \mathop{\limsup}_{\nu}\displaystyle\biggl\vert\frac{n_{\nu}}{N_{\nu}} - f\biggl\vert = \order{1}, \text{  with $P_{0}-$probability $1$.}
        \end{equation*}
\end{description}
We note that Conditions~\nameref{consistency} - \nameref{fraction} are necessary to produce consistency of the sample-weighted pseudo-posterior estimator and, by extension the MLE of the sample-weighted likelihood,
\begin{equation*}
        \mathbb{P}^{\pi}_{N_{\nu}}\ell_{\hat{\theta}_{\pi,N_{\nu}}} \geq \mathop{\sup}_{\theta,\nu}\mathbb{P}^{\pi}_{N_{\nu}}\ell_{\theta} - \smallO_{P_{\theta_{0}},P_{\nu}}\left(N_{\nu}^{-1}\right)
\end{equation*}
and $\hat{\theta}_{\pi,N_{\nu}} \ipnu \theta_{0}$, which was previously shown by \citet{2015arXiv150707050S, 2018dep}.  This paper focuses on achieving correct uncertainty quantification from the pseudo-posterior by adjusting credibility sets achieved from Bayesian hierarchical model specifications that are guaranteed to asymptotically contract on correct frequentist confidence intervals, which will rely or build upon this consistency result.  Bounding the supremum of over all $\nu$ of the inverse of inclusion probabilities from above in Condition~\nameref{bounded} is equivalent to bounding \emph{all} of the inclusion probabilities away from $0$.  This assumption is used by \citet{2015arXiv150707050S, 2018dep} to achieve their consistency result by ensuring that no portion of the population is systemically excluded from the sample such that the portion may never be sampled, which could otherwise lead to unattenuating bias. This requirement is used in essentially all consistency results in the literature; see, for example, \citet{pfeffermann98}.  We note that Condition~\nameref{deprestrict} defines two sets containing pairs of population units.  The set $S_{\nu 1}$ contains units, such as those within clusters, where sampling dependence among units is asymptotically \emph{unattenuated}, so long as the \emph{number} of dependent pairs of units is $\mathcal{O}\left(N_{\nu}\right)$.   The set, $S_{\nu 2}$, by contrast, contains pairs of units (e.g., pairs of units where each unit of the pair resides in a \emph{different} cluster or PSU from the other) where dependence \emph{is} required to asymptotically attenuate to $0$.  This condition relaxes the usual assumption of asymptotic independence among \emph{all} units (e.g., requiring that all units are in $S_{\nu 2}$) that has been typically used to guarantee the consistency result for $\hat{\theta}_{\pi,N_{\nu}} \ipnu \theta_{0}$.  While the more restrictive condition is met by nearly all single-stage designs used, in practice (including SRS), it does not apply to multistage clustered sampling designs (which our Condition~\nameref{deprestrict} \emph{does} cover).  For example, in a 2-stage sampling design where the first stage conducts a sampling of clusters of units, the set $S_{\nu 1}$ captures pairs of units \emph{within} PSU, whose dependence will not asymptotically attenuate to $0$, while $S_{\nu 2}$ contains pairs of units \emph{between} PSUs, where asymptotic independence is required; that is, the PSUs are required to be asymptotically independent.  See \citet{2015arXiv150707050S, 2018dep} for extensive details and proofs.
Lastly, Condition~\nameref{fraction} is utilized by \citet{2015arXiv150707050S, 2018dep} and relaxes the assumption of an asymptotically $0$ sampling fraction used in \citet{10.2307/1403631}.  \citet{10.2307/1403631} assume an asymptotically $0$ sampling fraction in order to approximate the variance of their statistic of interest with respect to the joint distribution, $(P_{\nu},P_{\theta_{0}})$, with just the marginal population generating distribution, $P_{\theta_{0}}$; in other words, they \emph{ignore} the sampling design distribution, $P_{\nu}$.  \citet{2015arXiv150707050S} show on page $27$ that Condition~\nameref{fraction} allows the replacement of $f \times N_{\nu}$ with $n_{\nu}$ for $\nu$ sufficiently large to specify the rate of contraction for their consistency result.  Their results also go through in the special case that the sampling fraction asymptotes to exactly $0$.

\begin{theorem}[Asymptotic Normality of the Pseudo-MLE]
\label{explimit}
Suppose conditions ~\nameref{continuity}-\nameref{fraction} hold.  Then
\begin{align}\label{limit}
\sqrt{N_{\nu}}\left(\hat{\theta}_{\pi,N_{\nu}}-\theta_{0}\right) &= - H_{\theta_{0}}^{-1}\frac{1}{\sqrt{N_{\nu}}}\mathop{\sum}_{i=1}^{N_{\nu}}\frac{\delta_{\nu i}}{\pi_{\nu i}}\dot{\ell}_{\theta_{0}}(\mbf{X}_{\nu i}) + \smallO_{P_{\theta_{0}},P_{\nu}}(1)\\
& = - H_{\theta_{0}}^{-1}\sqrt{N_{\nu}}\mathbb{P}^{\pi}_{N_{\nu}}\dot{\ell}_{\theta_{0}} + \smallO_{P_{\theta_{0}},P_{\nu}}(1)\\
& = - H_{\theta_{0}}^{-1}\mathbb{G}^{\pi}_{N_{\nu}}\dot{\ell}_{\theta_{0}} + \smallO_{P_{\theta_{0}},P_{\nu}}(1).
\end{align}
\end{theorem}

\begin{theorem}[Aysmptotic Variance of the Pseudo-MLE]
\label{pseudo-mle}
Suppose conditions ~\nameref{continuity}-\nameref{bounded} hold.  Then
\begin{subequations}\label{varpseudomle}
\begin{align}
\mbox{Var}_{P_{\theta_{0}},P_{\nu}} \{ -H_{\theta_{0}}^{-1}\sqrt{N_{\nu}}\mathbb{P}^{\pi}_{N_{\nu}}\dot{\ell}_{\theta_{0}} \} & = H_{\theta_{0}}^{-1}J^{\pi}_{\theta_{0}}H_{\theta_{0}}^{-1}\\
&= H_{\theta_{0}}^{-1}\left[J_{\theta_{0}} + \frac{1}{N_{\nu}}\mathop{\sum}_{i=1}^{N_{\nu}}\mathbb{E}_{P_{\theta_{0}}}\left\{\left[\frac{1}{\pi_{\nu i}}-1\right]\dot{\ell}_{\theta_{0}}(\mbf{X}_{\nu i})\dot{\ell}_{\theta_{0}}(\mbf{X}_{\nu i})^{T}\right\}\right]H_{\theta_{0}}^{-1}\label{warpMLE}\\
&\leq \gamma H_{\theta_{0}}^{-1}J_{\theta_{0}}H_{\theta_{0}}^{-1} = \gamma H_{\theta_{0}}^{-1}.\label{scaleMLE}
\end{align}
\end{subequations}
\end{theorem}
The $\gamma$ upper bound in Equation~\ref{scaleMLE} demonstrates a multiplicative injury to $\sqrt{N_{\nu}}$ convergence rate achieved for the MLE (under simple random sample of size, $N_{\nu}$, the population size) in the case of the pseudo-MLE.   The larger is $\gamma$, the more varied will be information in the samples around that for the population, which indicates a decreasing efficiency of the sampling design.   The amount of injury would be higher for less efficient sampling designs. The maximum penalty paid is a uniformly inflated scale, which will produce wider confidence regions.  Theorem \ref{pseudo-mle} does not restrict the possibility that some designs may be \emph{more} efficient than an SRS or that the efficiency varies by model parameter.
Equation~\ref{warpMLE} demonstrates that the shape or geometry of the limiting distribution will be impacted in the case of unequal sampling inclusion probabilities.  This ``warping" effect would be expected to be more pronounced in a highly-skewed proportion-to-size sampling design than in an unequally-weighted stratified sampling design with relatively few strata.  We demonstrate both of these (warping and scaling) phenomena via simulations in Section \ref{sec:sims}.

\begin{theorem}[Asymptotic Distribution of the Pseudo-Posterior]
\label{pseudo-posterior}
Suppose conditions ~\nameref{continuity}-\nameref{fraction} hold.  Then
\begin{equation}
\mathop{\sup}_{B\in\Theta}\bigg\vert \Pi^{\pi}_{N_{\nu}}\left(\theta \in B \mid\mbf{X}_{\nu}\bm{\delta}_{\nu}\right) - \mathcal{N}_{\hat{\theta}_{\pi,N_{\nu}},N_{\nu}^{-1}H_{\theta_{0}}^{-1}}\left(B\right)\bigg\vert \mathop{\rightarrow}^{P_{\theta_{0}},P_{\nu}} 0,
\label{pseudopost}
\end{equation}
where $\hat{\theta}_{\pi,N_{\nu}}$ may be the pseudo-MLE or the pseudo-posterior mean.
\end{theorem}

The different forms of the asymptotic covariance matrices for the pseudo-MLE (Theorem~\ref{pseudo-mle}), on the one hand, and the pseudo-posterior (Theorem~\ref{pseudo-posterior}), on the other hand, are driven by the failure of Bartlett's second identity under informative sampling $J^{\pi}_{\theta_{0}} \ne J_{\theta_{0}} =  H_{\theta_{0}}$.   This difference motivates our post-processing step, which we next introduce, that performs multiplicative adjustments to draws from the pseudo-posterior distribution such that their covariance is approximately equal to that of the pseudo-MLE.  Please see Section~\ref{app:proofs} of the Appendix for detailed proofs of Theorems~$1-3$.

\section{Post-processing the pseudo-posterior }\label{sec:adjustment}
From Section \ref{asymptotics} we see that the asymptotic covariance of the pseudo-MLE or pseudo-posterior mean is $H_{\theta_{0}}^{-1} J^{\pi}_{\theta_{0}} H_{\theta_{0}}^{-1}$, yet the asymptotic covariance of our samples drawn from the pseudo-posterior is $H_{\theta_{0}}^{-1}$. This is analogous to the differences observed in \citet{2009arXiv0911.5357R}, though our formulation for $J^{\pi}_{\theta_{0}}$ (and also $H^{\pi}_{\theta_{0}}$) arises from a random-sampling mechanism, which we leverage in the sequel to perform a post-hoc adjustment to draws from the pseudo-posterior.
Let  $\hat{\theta}_m$ represent the sample from the pseudo-posterior for $m = 1, \ldots, M$ draws with sample mean $\bar{\theta}$.
Define the adjusted sample:
\begin{equation}\label{eq:adjustment}
\hat{\theta}^{a}_m = \left(\hat{\theta}_m - \bar{\theta}\right) R^{-1}_2 R_1  + \bar{\theta},
\end{equation}
where $R'_1 R_1 = H_{\theta_{0}}^{-1} J^{\pi}_{\theta_{0}} H_{\theta_{0}}^{-1}$ and $R'_2 R_2 = H_{\theta_{0}}^{-1}$.
We may loosely think of $R^{-1}_2 R_1$ as a multivariate `design effect' adjustment
(For the SRS sample, we expect Barlett's second identity to hold and thus $H_{\theta_{0}}^{-1} J_{\theta_{0}} H_{\theta_{0}}^{-1} = H_{\theta_{0}}^{-1}$ which is the same asymptotic variance as the unadjusted pseudo-posterior).
Since $\hat{\theta}_m \appsim N(\theta_{0}, N^{-1}_{\nu}H_{\theta_{0}}^{-1})$, we now have $\hat{\theta}^{a}_m \appsim N(\theta_{0}, N^{-1}_{\nu}H_{\theta_{0}}^{-1} J^{\pi}_{\theta_{0}} H_{\theta_{0}}^{-1})$, which is the asymptotic distribution of the MLE under the pseudo-likelihood.
Unlike \citet{2009arXiv0911.5357R}, who pre-compute the MLE and change the geometry of their posterior sampler, our implementation is applied as a post-hoc projection of the pseudo-posterior sample, leaving the initial Monte Carlo sampler intact.  So the data analyst may use the Monte Carlo sampler that they designed for population model estimation (under simple random sampling).

For composite likelihoods, \citet{2009arXiv0911.5357R} calculate $\mbox{Var}_{P_{\theta_{0}}}\dot{\ell}_{\theta_{0}} =  J_{\theta_{0}}$ analytically. However, we have an additional distribution $P_{\nu}$ for the sampling design which is unlikely to be in analytic form. In practice, the design is often algorithmically defined; for example designs may use the sorting and clustering of population units in addition to unequal probabilities of selection. Rather than assuming a simplifying model for this distribution, we instead approximate the joint distribution $(P_{\theta_{0}}, P_{\nu})$ with the empirical distribution by resampling the units and associated response values.

Under a multistage sampling design with primary sampling units (PSUs) constructed as blocks (e.g., geographic regions or households) of last stage units (e.g., persons), we would re-sample a subset of the PSUs that contain dependent last-stage units, followed by including all last-stage units with each PSU.
We use information about the PSU memberships of each last stage unit in the observed sample in order to conduct the resampling.   The data analyst is expected to have this information about the structure of the sampling design, in addition to possessing the sampling weights for the last stage units (e.g., persons).  It is necessary when conducting the resampling to explicitly re-sample blocks of units, such as PSUs, when member units express dependence.
Such a procedure preserves the dependence structure within the replicate re-samples. This resampling procedure ensures our adjustment properly estimates the scale inflation of the pseudo-posterior distribution induced by the dependent step(s).   We use a simple random sampling without replacement (SRSWOR) procedure to re-sample the PSUs because they are nearly independent from one another, in practice.
Equivalently, the survey producer may issue sets of replicates weights which are created internally by using strata and cluster information. The data analyst can then skip directly to the estimation with the provided replicates (Step \ref{alg:repsum}). This variance estimation approach is a hybrid because it uses the Taylor linear expansion
to create transformed variables
\[(\hat{\psi} - \psi_0) = H_{\theta_0} (\hat{\theta} - \theta_0) \approx \mathop{\sum}_{i \in S} w_i \dot{\ell}_{\hat{\theta}}(\mbf{X}_{i}) = \mathop{\sum}_{i \in S} w_i z_i (\hat{\theta}).
\]
Variance estimations methods (Taylor linearization or replication methods) are then applied to $\mathop{\sum}_{i \in S} w_i \dot{\ell}_{\hat{\theta}}(\mbf{X}_{i})$ where the `total' is the estimate, and $\hat{\theta}$ is a plug-in, calculated only once.

Algorithm \ref{algo:R2R1} provides a simple and computationally efficient resampling approach to estimate $\mbox{Var}_{P_{\theta_{0}},P_{\nu}}\left[ \mathbb{P}^{\pi}_{N_{\nu}}\dot{\ell}_{\theta_{0}} \right] =  J^{\pi}_{\theta_{0}}$.
We recall from Section \ref{prelim} that $H_{\theta_{0}} = -\mathbb{E}_{P_{\theta_{0}}}\ddot{\ell}_{\theta_{0}}$ and $ H^{\pi}_{\theta_{0}} =-\mathbb{E}_{P_{\theta_{0}},P_{\nu}}\left[\mathbb{P}^{\pi}_{N_{\nu}}\ddot{\ell}_{\theta_{0}}\right] = H_{\theta_{0}}$.
Therefore, consistent estimates of $H_{\theta_{0}}$ are available without Algorithm \ref{algo:R2R1}. Both the plug-in estimate $-\mathop{\sum}_{i \in S} w_i \ddot{\ell}_{\bar{\theta}}(\mbf{X}_{i})$ and the posterior average $-\frac{1}{M}\mathop{\sum}_{m=1}^{M} \mathop{\sum}_{i \in S} w_i \ddot{\ell}_{\hat{\theta}_{m}}(\mbf{X}_{i})$ using the original sample $S$ will provide consistent estimates of $H_{\theta_{0}}$. (We drop the ``$\nu$" subscript from $\mbf{X}$ for readability). In our R implementation (Appendix \ref{app:code}), we use the plug-in estimate. Estimating $\hat{H}_{\theta_{0}}$ within each replication in Algorithm \ref{algo:R2R1} is also possible: $\hat{H}_{\theta_{0}} = \frac{1}{R}\mathop{\sum}_{r=1}^{R} h_r$ with $h_r = \mathop{\sum}_{l \in S^{r}} \tilde{w}^{r}_l \ddot{\ell}_{\bar{\theta}}(\mbf{X}^{r}_{l})$. However, the estimation of $\hat{J}^{\pi}_{\theta_{0}}$ cannot be performed without estimating across-PSU (or across-replicate) variance.
For simplicity, we use half the PSUs from the sample in each replicate \citep{Preston09}. Other resampling without replacement approaches should be effective \citep{Rao92}. However, sampling the PSUs \emph{with} replacement under-estimates the variance when the number of PSUs nested within strata is very small because with replacement sampling inaccurately reproduces the sampling design of PSUs from the population. For example, the NSDUH sample only has two PSUs available per strata.

\IncMargin{1em}
\begin{algorithm}
\DontPrintSemicolon
\SetKwInOut{Input}{input}\SetKwInOut{Output}{output}
\Input{
$\hat{\theta}_m$ from the pseudo-posterior \eqref{inform_post}\\
$\{j,k\}$ indicators for PSUs $j = 1,\ldots, J_{k}$ and Strata $k = 1,\ldots, K$. \\
$\{w_{ijk}, \mbf{X}_{ijk}\}$ for all $i$ in $1,\ldots, I_{jk}$ for every $\{j,k\}$\\
$R$ number of replicates.
}
\Output{Adjusted sample $\hat{\theta}^{a}_m$}
\BlankLine
Calculate the posterior mean $\bar{\theta} = \frac{1}{M}\mathop{\sum}_{m=1}^{M}\hat{\theta}_m$\;
Calculate plug-in $\hat{H}_{\theta_{0}}  =-\mathop{\sum}_{i \in S} w_i \ddot{\ell}_{\bar{\theta}}(\mbf{X}_{i})$\;
\For{\emph{Replicates} $r\leftarrow 1$ \KwTo $R$}{
\emph{Subsample PSUs without replacement (SRSWOR)}\;
\For{\emph{Strata} $k\leftarrow 1$ \KwTo $K$}{
	\emph{Sample half the PSUs within strata $k$}:
	$\{j'\}^{r}_{k}$ with $|\{j'\}^{r}_{k}| = J_{k}/2$ \;
	Take all units within each selected PSU: $S^{r}_{k} = \cup_{j'} \cup_{i}\{ij'k\}$\;
	Define sample $\{w^{r}_{l}, \mbf{X}^{r}_{l}\}$ for new index $l \in S^{r}_{k}$\;
	Double weights $\hat{w}_{l}^{r} = 2 w^{r}_{l}$ \;
}
Combine samples across strata: $S^{r} = \cup_k S^{r}_{k}$\;
Normalize weights  $\tilde{w}^{r}_l = \hat{w}^{r}_{l} \left(n / \mathop{\sum}_{l \in S^{r}} \hat{w}^{r}_l \right)$\;
	Evaluate $j_r = \mathop{\sum}_{l \in S^{r}} \tilde{w}^{r}_l \dot{\ell}_{\bar{\theta}}(\mbf{X}^{r}_{l})$\; \label{alg:repsum}
}
Calculate 	$\hat{J}^{\pi}_{\theta_{0}} = \frac{1}{R-1}\mathop{\sum}_{r=1}^{R}(j_r - \bar{j})(j_r - \bar{j})^t$ with $\bar{j} = \frac{1}{R}\mathop{\sum}_{r=1}^{R} j_r$\;
Calculate  $\hat{R}_1$ via Cholesky decomposition: $\hat{R}'_1 \hat{R}_1 = \hat{H}_{\theta_{0}}^{-1} \hat{J}^{\pi}_{\theta_{0}} \hat{H}_{\theta_{0}}^{-1}$\;
Calculate $\hat{R}_2$ via Cholesky decomposition: $\hat{R}'_2 \hat{R}_2 = \hat{H}_{\theta_{0}}^{-1}$\;
Calculate inverse $\hat{R}^{-1}_2$\;
Evaluate Eq. \ref{eq:adjustment}: $\hat{\theta}^{a}_m = \left(\hat{\theta}_m - \bar{\theta}\right) \hat{R}^{-1}_2 \hat{R}_1  + \bar{\theta}$\;
\caption{Adjust pseudo-posterior to correct for complex survey design}
\label{algo:R2R1}
\end{algorithm}\DecMargin{1em}
\FloatBarrier
\section{Simulation Study}\label{sec:sims}
We construct a population model to address our inferential interest of a binary outcome $y$ with a linear predictor $\mu$.
\begin{equation} \label{pop_like}
y_{i} \mid \mu_{i} \ind Bern \left(F_l(\mu_{i}) \right),~ i = 1,\ldots,N
\end{equation}
where $F_l$ is the cumulative distribution function for the logistic distribution.
The first set of simulations (Section \ref{sec:DE}) is based on equal probability sampling. We let $\mu$ depend on a single predictor $x_1$.
The second set of simulations (Section \ref{sec:PPS1}) is based on unequal probability sampling. We let $\mu$ depend on two predictors $x_1$ and $x_2$, where $x_2$ is a size variable to set the selection probabilities into the sample.
The third set of simulations (Section \ref{sec:PPS3}) is also based on unequal probability sampling, but we let $\mu$ depend on three predictors $x_1$, $x_2$, and $z_2$, where the latter is a random cluster effect at the PSU level.  The quantity of inferential interest for all of our simulations is the estimation of the population model coefficients (intercept and slope) for $x_1$. The $(x_2,z_2)$ are nuisance.

The variable $x_1$ represents the observed information available for analysis, whereas $x_2$ represents auxiliary information available for setting inclusion probabilities used to conduct sampling, which is either ignored or not available for analysis. The $x_1$ and $x_2$ distributions are $\mathcal{N}(0,1)$ and $\mathcal{E}(r =1/5)$ with rate $r$, where $\mathcal{N}(\cdot)$ and $\mathcal{E}(\cdot)$ represent normal and exponential distributions, respectively. The cluster effect $z_2$ is neither a design variable used for sampling nor part of the analytical model, but is a nuisance representing unknown and un-modeled dependence between units within the same cluster (PSU). We choose $z_2 \sim \mathcal{E}(1/5)$ for a skewed distribution.

We formulate the logarithm of the sampling-weighted pseudo-likelihood for estimating $(\mu,\theta)$ from our observed data for the $ n\leq N$ sampled units,
\begin{align}\label{pseudo_like}
\log\left[\mathop{\prod}_{i=1}^{n} p\left(y_{i}\mid x_{1i}, \beta_0, \beta_1 \right)^{\tilde{w}_{i}}\right] &= \mathop{\sum}_{i=1}^{n}\tilde{w}_{i}\log p \left( y_{i}\mid x_{1i}, \beta_0, \beta_1\right) \nonumber\\
&= \mathop{\sum}_{i=1}^{n} \tilde{w}_{i} y_{i} \log(F_l( \beta_0 + x_{1i} \beta_1)) \\
& \quad+ \tilde{w}_{i} (1- y_{i}) \log(1-F_l( \beta_0 + x_{1i} \beta_1)), \nonumber
\end{align}
where $\theta = (\beta_0, \beta_1)$, $\mu_i =   \beta_0 + x_{1i} \beta_1$, and the sampling weights, $\tilde{w}_{i}$ are normalized such that the sum of the weights equals the sample size $\mathop{\sum}_{i=1}^{n}\tilde{w}_{i} = n$.

Finally, we estimate the joint posterior distribution using Equation~\ref{pseudo_like}, coupled with our prior distribution assignments, using the NUTS Hamiltonian Monte Carlo algorithm implemented in Stan \citep{stan:2015, Rstan}. All computations were performed in R \citep{R}. See Appendix \ref{app:code} for example code for fitting and adjusting a Stan model. Generating samples from the different sample designs can be implemented in several ways. We chose to use the `sampling' package \citep{Rsampling} for the PPS method of \cite{BrewerPPS} and for systematic sampling.

\subsection{Simulation Designs}
In the following subsections we discuss how we construct sampling design distributions, $P_{\nu}$, that will induce dependence and skewed information about the population in the observed sample as a means of assessing the performance of our post-processing adjustment procedure specified in Algorithm \ref{algo:R2R1}. In Section \ref{sec:results}, we will assess whether the adjustments performed to the posterior draws generate credibility sets that achieve nominal frequentist coverage.  We recall from Section \ref{sec:intro} that the survey sampling literature defines the design effect (DEFF) as the ratio of the variance of a estimate for the population mean $\bar{Y}$ under a complex survey design compared to the variance under simple random sampling: $\mbox{DEFF}_{\bar{Y}} = \Var_{P_{\nu}}(\widehat{\bar{Y}}) / \Var_{\mbox{SRS}}(\widehat{\bar{Y}})$.
In addition to nominal coverage, we are also interested in comparing our model-based design effects to the standard $\mbox{DEFF}_{\bar{Y}}$ output of design-based survey software, such as the R `survey' package \citep{Rsurvey} which implements survey-weighted maximum likelihood for point estimation and by default uses Taylor linearization methods for variance estimation. We estimate the marginal design effect for each parameter:  $\mbox{DEFF}_{\theta} = \mbox{diag}\{H_{\theta}^{-1} J^{\pi}_{\theta} H_{\theta}^{-1}\}/\mbox{diag}\{H_{\theta}^{-1}\}$. These parameter-specific $\mbox{DEFF}_{\theta}$ provide an estimate of the marginal rescaling induced by the complex sample design relative to a simple random sample.

\subsubsection{Equal Probability Dependent Designs (DE)}\label{sec:DE}
For these designs, we induce dependence in the observed samples by clustering units; for example, by aggregating individuals in the population by geographically-indexed domains.   This type of clustering or grouping of units is performed by the sampling designers, in practice, in order to control the costs (in this case, travel and labor costs) of administering the survey.   It is typically the case that the clustering structure will be coincident with a dependence structure in the population variables of interest; for example, geographically-indexed domains capture a spatial dependence among measures for individuals induced by similarities in culture and economic factors.   The effect is that individuals are sampled in dependent groups or clusters, which is expected to lower the amount of information about the population in a realized random sample under this type of sampling design as compared to a simple random sampling of individuals taken from the same population. Even if a sampling design distribution, $P_{\nu}$, is not informative, the design will induce a scale inflation in the asymptotic covariance of the posterior distribution if the design includes a stage that samples dependent clusters.   Our theoretical results don't directly address this possibility but instead focus on warping and scale adjustments due to approximation error of the pseudo-posterior induced by unequal weighting.  However, we demonstrate in the sequel that the post-processing adjustment procedure of Algorithm 1, nevertheless, adjusts the scale of the posterior distribution under this scenario to achieve nominal coverage.

The population generating model is
\[
\mu_i = 0.0 + 1.0 x_{1i}
\]
where the intercept was chosen such that the median of $\mu$ is 0, therefore the median of $F_l(\mu)$ is 0.5.

The first design (DE1), is a one-stage cluster design where clusters of size 5 are selected according to simple random sampling (SRS). All individuals have responses that are unconditionally independent. In other words, the clustering membership is randomized and uninformative.  Under this scenario, the pseudo-likelihood reduces to the true likelihood with correctly specified independence between units. Therefore, both the unadjusted MCMC samples $\hat{\theta}_m$ and the adjusted MCMC samples $\hat{\theta}^{a}_m$ should ideally have similar coverage.

The second design (DE5) is also a one-stage SRS design, except that all 5 members of each cluster have complete dependence. Both the $y$ and the $x_1$ have identical values within each cluster: $y_{ij} = y_{i'j}$ and $x_{1ij} = x_{1i'j}$ for all individuals $i \ne i'$ in cluster $j$. Under this scenario, the pseudo-likelihood is again reduced to the simple likelihood. While the likelihood is correctly specified for any given individual, joint cluster dependence is mis-specified as independence.
Effectively, the sum of the (equal) weights should really sum to $n/5$ rather than $n$. 
Under this scenario, the unadjusted MCMC samples $\hat{\theta}_m$ should have intervals that are too narrow by a factor of $\sqrt{5}$ while the adjusted intervals for $\hat{\theta}^{a}_m$ should be longer and achieve the nominal coverage. This idealized example, in which within cluster dependence is both unspecified in the analyst's model and complete, demonstrates the \emph{sensitivity} of the posterior (and pseudo-posterior) to the mis-specification of the effective sample size $n$ and the \emph{robustness} of Algorithm \ref{algo:R2R1} to correct for this.

\subsubsection{One stage unequal probability designs (PPS1)}\label{sec:PPS1}
For these next designs, we have no dependence induced by the clustering of units. Instead, we use an informative design $P_{\nu}$ which uses information from the population to sample units with unequal probabilities of selection; for example, selecting larger businesses with higher probability than smaller businesses in the Current Employment Statistics (CES) survey, administrated by the U.S. Bureau of Labor Statistics (BLS). In practice, these designs control costs because large businesses contribute proportionately more to estimates for industry totals, such as total production or number of employees. Further refinements to the design, such as stratification of units into size classes, also create statistical efficiencies by reducing the possibility of extreme sample outcomes (such as selecting a sample composed entirely of small businesses). Our theoretical results directly address these informative designs which lead to warping and scale effects due to the approximation error of the pseudo-posterior induced by unequal weighting. We demonstrate that our post-processing adjustment via Algorithm 1 achieves nominal coverage under these informative sampling designs.

The population generating model is now
\[
\mu_i = -1.88 + 1.0 x_{1i}+ 0.5 x_{2i}
\]
where the intercept was chosen such that the median of $\mu$ is approximately 0, therefore the median of $F_l(\mu)$ is approximately 0.5. The size measure used for sample selection is $\tilde{x}_{2i} = x_{2i} - \min_i (x_{2i}) + 1$.

Even though the population response $y$ was simulated with $\mu = f(x_1,x_2)$, we estimate the marginal models at the population level for $\mu = f(x_1)$. This exclusion of $x_2$ is analogous to the situation in which an analyst does not have access to all the sample design information and ensures that our sampling design instantiates informativeness (where $y$ is correlated with the selection variable, $x_{2}$, that defines inclusion probabilities). In particular, we estimate the models under informative design scenarios and compare the population fitted models, $\mu = f(x_1)$, to those from the samples. The first unequally weighted design is a one-stage probability proportional to size design (PPS1), where probabilities of selection are proportional to the size measure $\pi_i \propto \tilde{x}_{2i}$. For the same population model we also create a stratified design (SPPS1).  We add this additional design because stratification is expected to \emph{improve} the efficiency of the sampling design as compared to SRS because it will - on average - produce samples that are more informationally representative of the population, such that $\mbox{DEFF}_{\theta}$ may be less than $1$.  We demonstrate the our scale adjustment adapts to more efficient, as well as less efficient, sampling designs.  The population is sorted by size measure $\tilde{x}_{2}$ and then placed into 10 strata. We then select $n/10$ units from each strata $k$ with $\pi_{ik} \propto \tilde{x}_{2ik}$.

\subsubsection{Three stage unequal probability designs (PPS3)}\label{sec:PPS3}
The last set of designs combines feature of the first two sets. In practice, multistage designs such as the NSDUH first select geographic PSUs (such as states, counties, census tracts, etc) in proportion to a measure of population size. This provides both cost savings (collecting data in geographic clusters) and statistical efficiencies (higher population areas represent more of the population total), especially when combined with geographic-based stratification (e.g. by state). The final stages for multistage surveys are often the household and individual. The effect is that individuals within each PSU cluster may likely have outcome measures related to others in their household and geographic cluster. The within PSU dependence and the unequal probabilities of selection will induce both a scale inflation and a warping in the asymptotic covariance of the posterior distribution. Our theoretical results don’t directly address the rescaling due to within cluster dependence; however, our post-processing adjustment procedure of Algorithm 1, nevertheless, adjusts both the scale and shape of the posterior distribution under this scenario to achieve nominal coverage.

The population generating model is now
\[
\mu_{ij} = -1.88 + 1.0 x_{1ij}+ 0.25 x_{2ij} + 0.25 \bm{z}_{2j}
\]
where $z_{2j} \sim \mathcal{E}(1/5)$ is the random effect for PSU $j$. The median of $\mu$ is still close to 0, and the median of $F_l(\mu)$ is still close to 0.5. The size measure used for sample selection is $\tilde{x}_{2i} = x_{2i} - \min_i (x_{2i}) + 1$. Compared to the population model for PPS1 and SPPS1, the relationship between $y$ and the size variable $x_2$ is weaker ($0.25$ vs. $0.50$). This is often the case for household surveys compared to establishment surveys, because the amount of information available to the sample designer is much greater for establishments than for households.

The next design is a three-stage PPS design (PPS3), analogous to a household survey in which a geographic area is selected as a PSU, followed by a household (HH) and an individual. We employ a simplified, but broadly representative, version of the design used for NSDUH where we first select the PSU based on the size $\tilde{x}_{2}$ aggregated up to the PSU level. We next select 5 out of 10 HHs within each PSU, where the HH's are sorted based on an aggregate size measure from $\tilde{x}_{2}$ and sampled systematically (i.e. every other one along the rank sorted list). Finally, 1 of 3 individuals are selected within each household in proportion to the individual size measure $\tilde{x}_{2}$. The nested sampling within PSU, the systematic sampling of HHs, and the mutually exclusive sampling of individuals within HHs creates a sampling dependence that does not attenuate (i.e. factor). See \citet{2018dep} for a richer discussion  of the sources of sampling dependence.

We include a PSU level random effect $z_{2j}$ to allow for the possibility of un-modeled population level dependence that coincides with the sample design induced dependence and together reduce the effective sample size. For example, geographic covariates such as state or census tract may be related to the outcome of interest, but like $x_{2}$ they are unavailable to the analyst of a public use file due to confidentiality protections. We expect the unadjusted MCMC sample $\hat{\theta}_m$ to undercover both due to the warping effect from unequal weighting and due to the over-estimation of the effective sample size from the nuisance PSU dependence. We expect the adjusted MCMC sample $\hat{\theta}^{a}_m$ to capture this dependence, leading to wider uncertainty intervals with closer to nominal coverage.

Lastly, we include a stratified version of the design (SPPS3) in which the aggregate size variable for the PSUs is used to sort the clusters into 10 strata, which are then sampled in a three stage design. Since the size variable $x_2$ has a weaker relationship with the outcome, the impact of stratification will be weaker for SPPS3 compared to SPPS1. This example is the closest to our motivating NSDUH design and provides insight into the performance of Algorithm \ref{algo:R2R1} when resampling PSUs nested with strata.

\subsection{Results}\label{sec:results}
Table \ref{tab:sims} provides a summary of results for $100$ Monte Carlo realizations for each of the 6 designs based on a target nominal coverage of 90\%. A separate population with a specific formulation for the mean of the linear predictor is generated for each of DE1, DE5, (PPS1,SPPS1) and (PPS3, SPPS3) scenarios in each Monte Carlo realization.  A separate sample is taken from the population for each sampling design (of the $6$ total sampling designs). Estimation of points and intervals were conducted for each sample.   Total sample sizes of $n = 200$ were used to explore performance for moderate sample sizes. \cite{2018dep} demonstrate good convergence for this sample size and similar model settings. In other words, the bias for the posterior mean and the MLE is negligible, so we can focus only on the coverage. While discussion of the mean squared error may be interesting, that is a property of the point estimate not the uncertainty distribution. Since the unadjusted and the adjusted MCMC samples have identical means and are very close to the MLE, the bias and the MSE of the three are essentially the same.

For each Monte Carlo realization, we create $R = 100$ replicates to perform an adjustment via Algorithm \ref{algo:R2R1}. We consider coverage estimates from 85\% to 95\% to be reasonably close to the nominal 90\% given the simulation noise from the $100$ realizations. Marginal coverage is assessed from the two-sided intervals from sample quantiles $(q_{05},q_{95})$. For simplicity, joint coverage is assessed by comparing the Mahalanobis distance $(\hat{\theta} - \theta)'Var(\hat{\theta})(\hat{\theta} - \theta)$ to the 90\% quantile of a $\chi^2_2$ distribution. Figure \ref{fig:simscatter} displays one realization from each of the design simulations before and after adjustment to visually demonstrate the rescaling and rotation (to undo warping from unequal probability informative sampling) of the adjustment. It also compares the two pseudo-posterior ellipses to those from the pseudo-MLE. This pseudo-MLE ellipse is equivalent to the joint region from numerical samples in \cite{Wang2017}.
Figure \ref{fig:simmarg} displays the marginal distributions, medians, and 90\% quantiles for the unadjusted and adjusted pseudo-posterior distribution. The reference lines display the MLE based median and 90\% intervals which are equivalent to the marginal regions from numerical samples in \cite{Wang2017}.

\subsubsection{Coverage}
\begin{figure}
\centering
\includegraphics[width = 0.95\textwidth,
		page = 1,clip = true, trim = 0.25in 0.25in 0in 0.in]{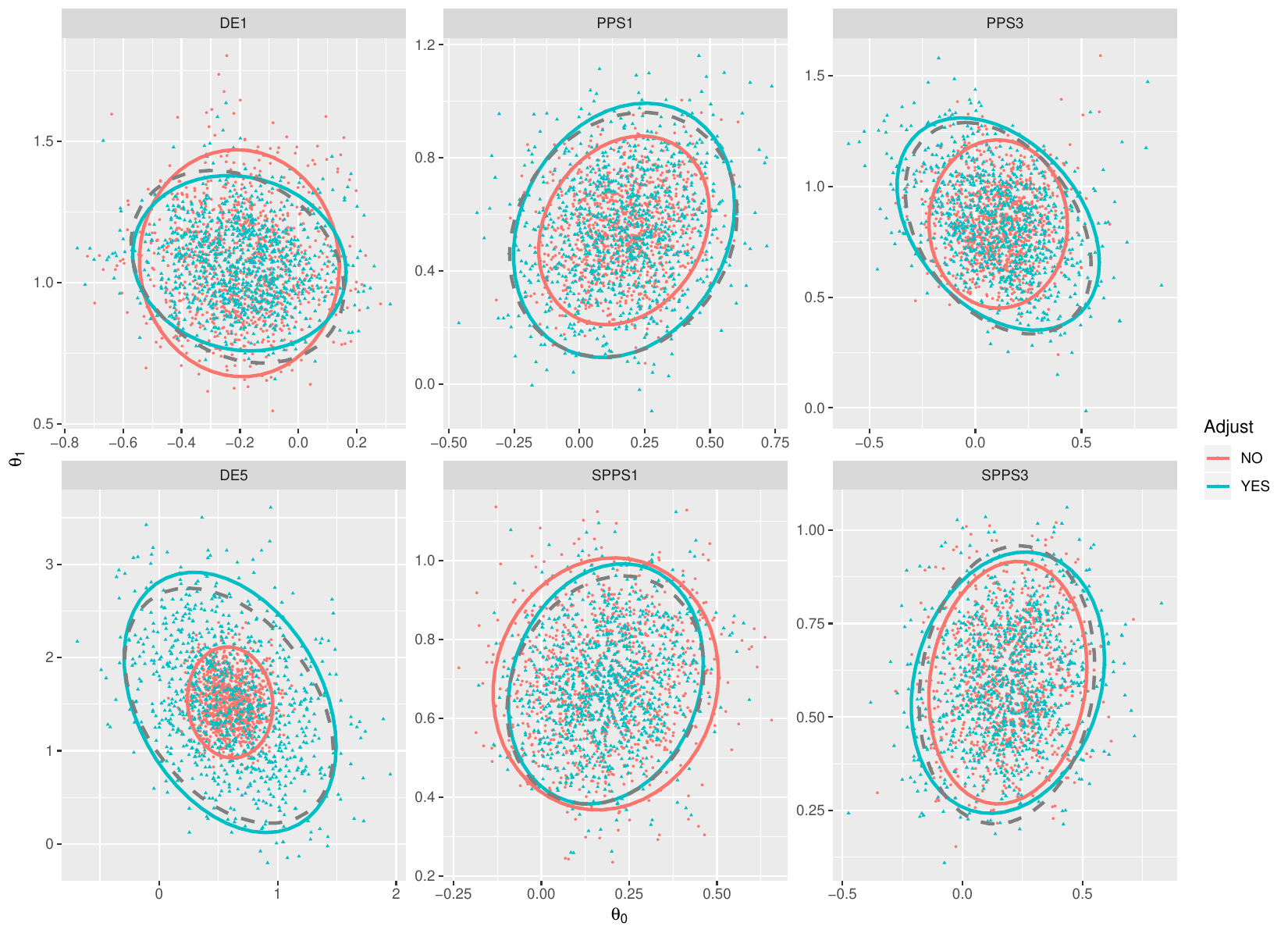}
\caption{Joint pseudo-posterior sample for the intercept (horizontal) and slope (vertical) for one realization of each of six sample designs. Unadjusted (red circles) and adjusted (blue triangles) with approximate 90\% density ellipses. Asymptotic normal 90\% ellipse for pseudo-MLE (dashed). Created with `ggplot2' \citep{Rggplot2}.}
\label{fig:simscatter}
\end{figure}

\begin{figure}
\centering
\includegraphics[width = 0.95\textwidth,
		page = 1,clip = true, trim = 0.25in 0.25in 0in 0.in]{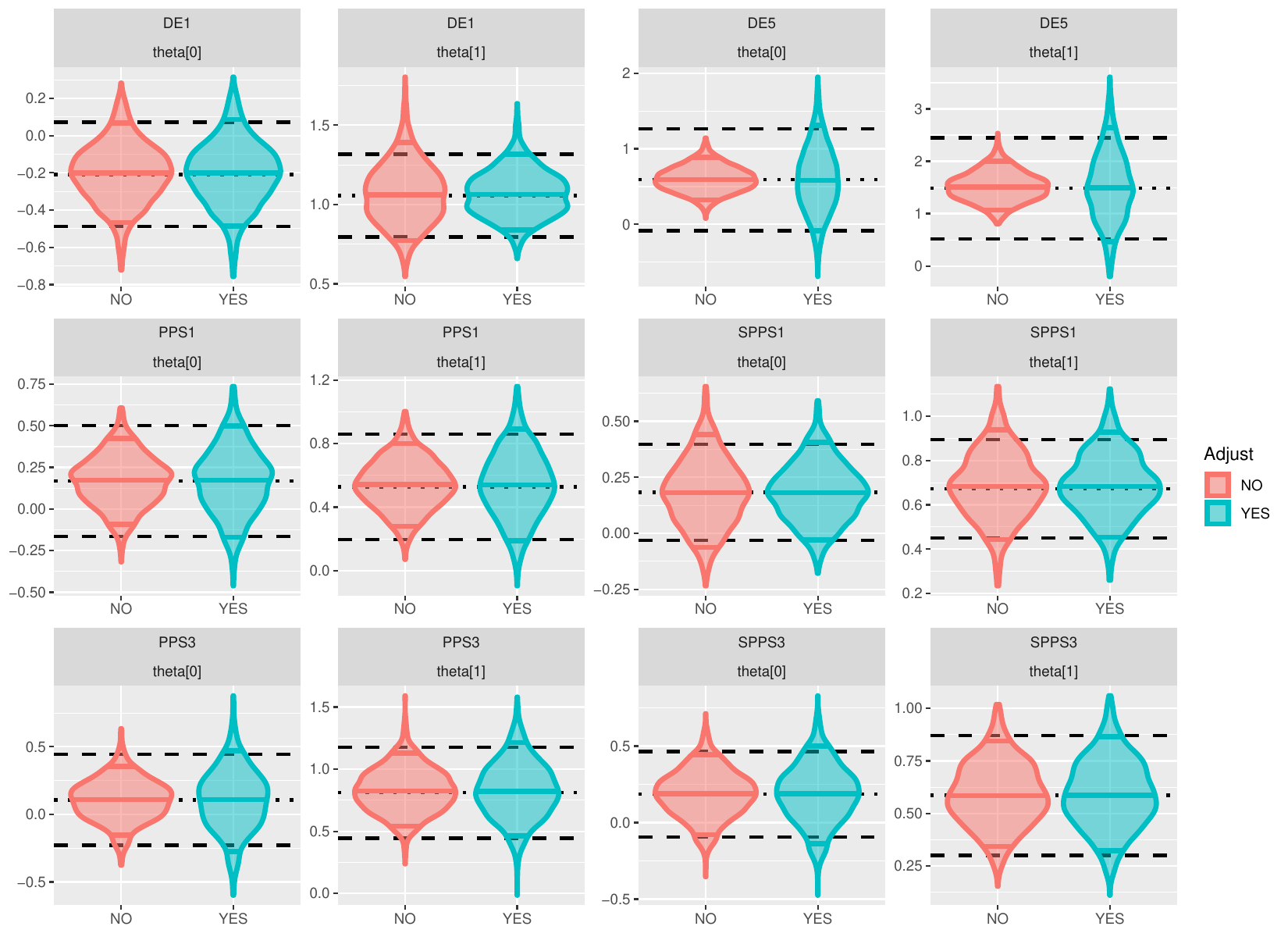}
\caption{Marginal pseudo-posterior sample for the intercept ($\theta_0$) and slope ($\theta_1$) for one realization of each of six sample designs. Unadjusted (left) and adjusted (right) with median and 90\% quantiles (solid bars). Asymptotic normal mean (dotted) and 90\% interval  (dashed) for pseudo-MLE. Created with `ggplot2' \citep{Rggplot2}.}
\label{fig:simmarg}
\end{figure}

The one-stage equal probability designs (DE1 and DE5) demonstrate that Algorithm \ref{algo:R2R1} is effective across the entire range of within cluster independence to complete cluster dependence. DE1 serves as a control, in which no adjustment should be needed. The marginal coverage and interval widths for the adjusted sample $\hat{\theta}^{a}_m$ are slightly lower than for the unadjusted $\hat{\theta}_m$ but the joint elliptical coverage is about as good. DE5 serves as an extreme example under which $\hat{\theta}_m$ is clearly undercovering and $\hat{\theta}^{a}_m$ performs much better. Figures~\ref{fig:simscatter} and \ref{fig:simmarg} show one realization in which the densities mostly overlap for DE1. For DE5, we see the adjusted density is much more diffused and indicates some design-induced dependence between the parameters. This may explain why the joint coverage for $\hat{\theta}_m$ is even worse than the marginal coverage and it suggests that a naive rescaling of the weights by 5 might not lead to correct joint coverage as postulated in section \ref{sec:DE}. Comparisons to the asymptotic MLE distribution reflect good but not perfect alignment between the adjusted distribution and the asymptotic distribution. We expect this because the adjusted pseudo-posterior, while having improved asymptotic coverage, still maintains its small sample properties. For example, the marginal distributions are not perfectly symmetric or unimodal.

The one-stage unequal probability designs (PPS1 and SPPS1) demonstrate the warping effect without the presence of within cluster dependence. PPS1 demonstrates improvements for both marginal and joint coverage. The stratified version (SPPS1) shows that the unadjusted sample $\hat{\theta}_m$ is over-covering, particularly for the joint region. For the moderate sample size $n= 200$, the adjusted coverage shows a decrease for the intercept but a much closer to nominal coverage for the joint region (88\% vs. 99\%).  Figures~\ref{fig:simscatter} and \ref{fig:simmarg} show a similar pattern. The increase in dispersion for the PPS1 design  is reduced and offset by stratification in SPPS1. These designs are similar to establishment surveys such as the CES, which may use frame data to form efficient strata and samples for businesses. Again, the adjusted pseudo-posterior compares well to the asymptotic MLE distribution, while still maintaining small sample properties.

The three-stage designs with PSU level dependence (PPS3 and SPPS3) show similar results. The unequal selection is weaker in the three stage designs than in the one-stage. Therefore the stratification does not lead to much gain in efficiency. Both designs show an improvement in coverage for both marginal and joint coverage and match well (but not identically) to the asymptotic MLE distribution. This is consistent with results from the one-stage designs, but combines the unequal weighting, within PSU dependence, and stratification into a single design, similar to household surveys such as the NSDUH.

\subsubsection{Design Effects}
We next compare the parameter-specific $\mbox{DEFF}_{\theta}$ to the $\mbox{DEFF}_{\bar{Y}}$ based on Taylor linearization \citep{lumley2004}. Table \ref{tab:sims} shows that the design effect for the intercept $\theta_0$ is very similar to the overall design effect for $y$, where the latter is computed from Taylor linearization methods. This is not surprising, since an intercept is very similar to a mean.
Examining the design effect for the slope $\theta_1$, we see that the effect of the design is typically less dramatic than for the intercept but still notably different from 1. We remind the reader that these estimates for design effects assume the bias has been removed due to incorporation of the weights and do not suggest that equally weighted likelihoods will lead to estimates for slopes that have correct coverage. For comparisons between consistent weighted estimates and biased unweighted estimates see \citet{2015arXiv150707050S, 2018dep}.

\begin{table}
\begin{center}
\caption{Summary of coverage, average width, and design effect estimates for simulations based on 90\% posterior intervals. Based on  $M = 100$ realizations with sample size $n = 200$, $R = 100$ replications, and population sizes $N = 5000, 5000, 6000$ for the SRS (DE1,DE5), one-stage PPS (PPS1,SPPS1), and and three-stage (PPS3, SPPS3) designs , respectively, where \emph{S} denotes the nesting within a stratified sampling stage.}
{
\resizebox{\columnwidth}{!}
{
\begin{tabular}{r|rr|rr|rr|rr|rr|rrr}
  \hline
Scenario &  \multicolumn{2}{c|}{Marginal $\theta_0$} & \multicolumn{2}{c|}{Marginal $\theta_1$} & \multicolumn{2}{c|}{Joint $\theta_0, \theta_1$}&
 \multicolumn{2}{c|}{Width $\theta_0$} &\multicolumn{2}{c|}{Width $\theta_1$} &
 \multicolumn{3}{c}{DEFF}\\
 & $\hat{\theta}_m$ & $\hat{\theta}^{a}_m$ & $\hat{\theta}_m$ & $\hat{\theta}^{a}_m$& $\hat{\theta}_m$ & $\hat{\theta}^{a}_m$&
  $\hat{\theta}_m$ & $\hat{\theta}^{a}_m$ & $\hat{\theta}_m$ & $\hat{\theta}^{a}_m$&
 $\theta_0$ & $\theta_1$ & $\bar{y}$\\
\hline
DE1&0.89 & 0.86 & 0.89 & 0.90 & 0.93& 0.87&
	  0.52 & 0.51& 0.64 & 0.63 &
	   0.96 & 0.99 & 0.97\\
DE5& 0.43& 0.81& 0.56& 0.94& 0.32& 0.88&
	   0.55& 1.24& 0.70& 1.60&
	   5.06& 5.26& 5.10\\ \hline
PPS1& 0.77& 0.88&0.83 &0.91 & 0.74& 0.93&
	   0.50& 0.69& 0.55&0.70 &
	   1.90& 1.63& 1.86\\
SPPS1& 0.91& 0.84& 0.96& 0.96& 0.99& 0.88&
	   0.49& 0.41& 0.54& 0.55 &
	   0.69& 1.02 & 0.71\\ \hline
PPS3& 0.74& 0.91& 0.79& 0.87& 0.75& 0.86&
	   0.51& 0.75& 0.57& 0.75&
	   2.20& 1.71& 2.13\\
SPPS3& 0.77& 0.95& 0.80& 0.87& 0.74& 0.87&
	   0.51& 0.73& 0.56& 0.71&
	   2.03& 1.59& 1.99\\
  \hline
\end{tabular}
}
\label{tab:sims}
}
\end{center}
\end{table}

\FloatBarrier

\section{Application: NSDUH}\label{sec:NSDUH}
\begin{figure}
\centering
\includegraphics[width = 0.60\textwidth,
		page = 1,clip = true, trim = 0.25in 0.25in 0in 0.in]{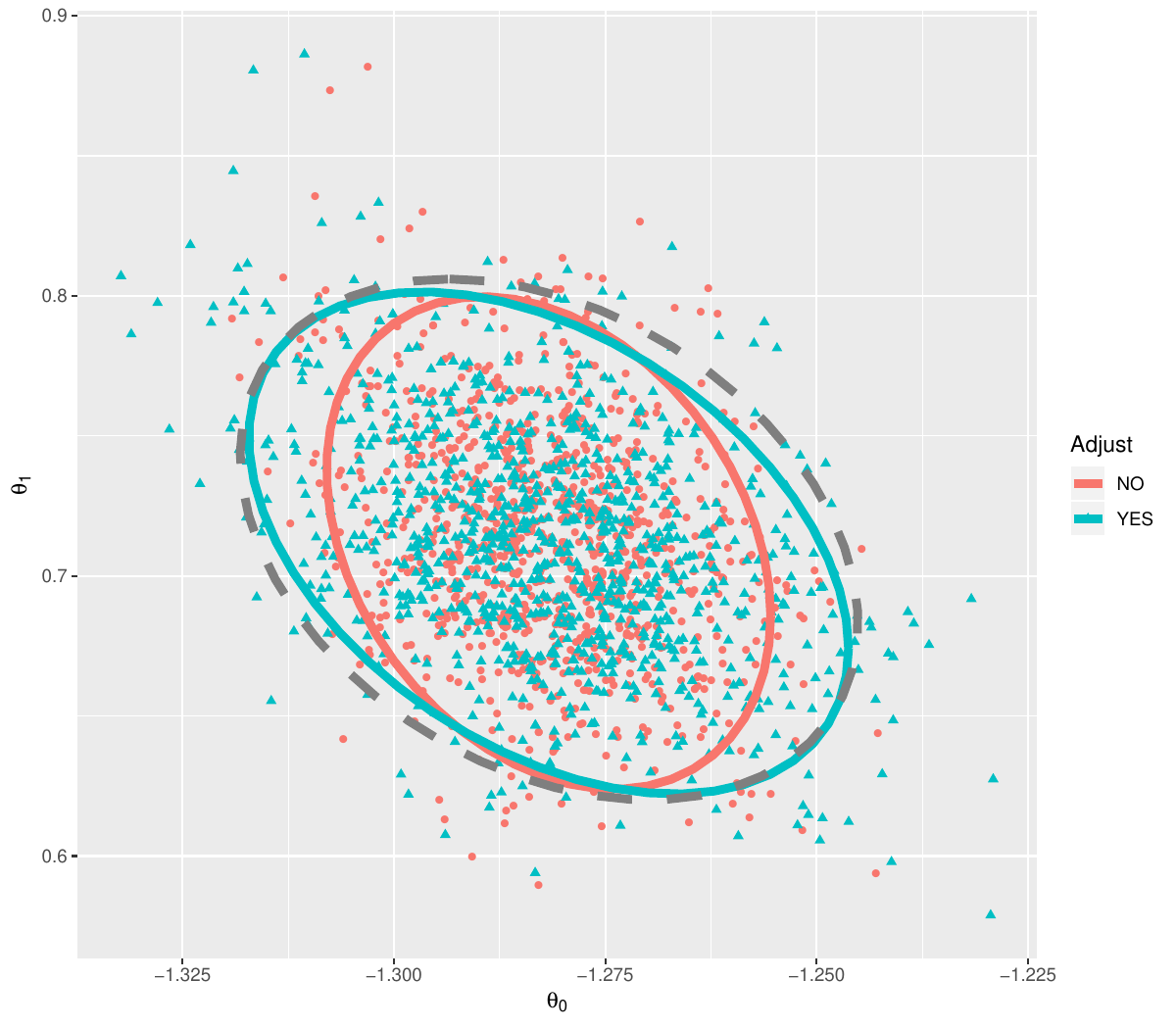}
\caption{Joint pseudo-posterior sample for the intercept (horizontal) and slope (vertical) for a logistic regression modeling current cigarette smoking by past year major depressive episode based the 2014 National Survey on Drug Use and Health. Unadjusted (red circles) and adjusted (blue triangles) draws with approximate 90\% density ellipses. Asymptotic normal 90\% ellipse for pseudo-MLE (dashed). Created with `ggplot2' \citep{Rggplot2}.}
\label{fig:NSDUHscatter}
\end{figure}

A simple logistic model relating current (past month) smoking status to the presence of a past year major depressive episode (MDE) was fit via the survey-weighted pseudo-posterior as described in Section~\ref{sec:sims} using probability-based analysis weights for adults from the 2014 NSDUH public use data set.
Overall the design effect for $\bar{y}$ from the is 1.87, and the parameter specific design effects are 1.88 for the intercept and 1.12 for the slope. In addition to the marginal rescaling, Figure \ref{fig:NSDUHscatter} demonstrates the presence of a joint warping effect from the complex sample design of the NSDUH.

The rates for both smoking and depression vary by age, urban/rural status, education, and other demographics. Some of these factors are related to the sample inclusion probabilities, thus weighting is needed to mitigate bias. For example see \cite{2018dep} for a comparison of the weighted and unweighted estimates. Some of these features (geographic and household) also correspond to nested clusters of sampling, creating the potential for intercluster correlations. Together these design features contributed to an almost doubling the of the variance associated with the mean and intercept, but a relatively smaller increase in the variance associated with the slope.

\begin{figure}
\centering
\includegraphics[width = 0.60\textwidth,
		page = 1,clip = true, trim = 0.25in 0.25in 0in 0.in]{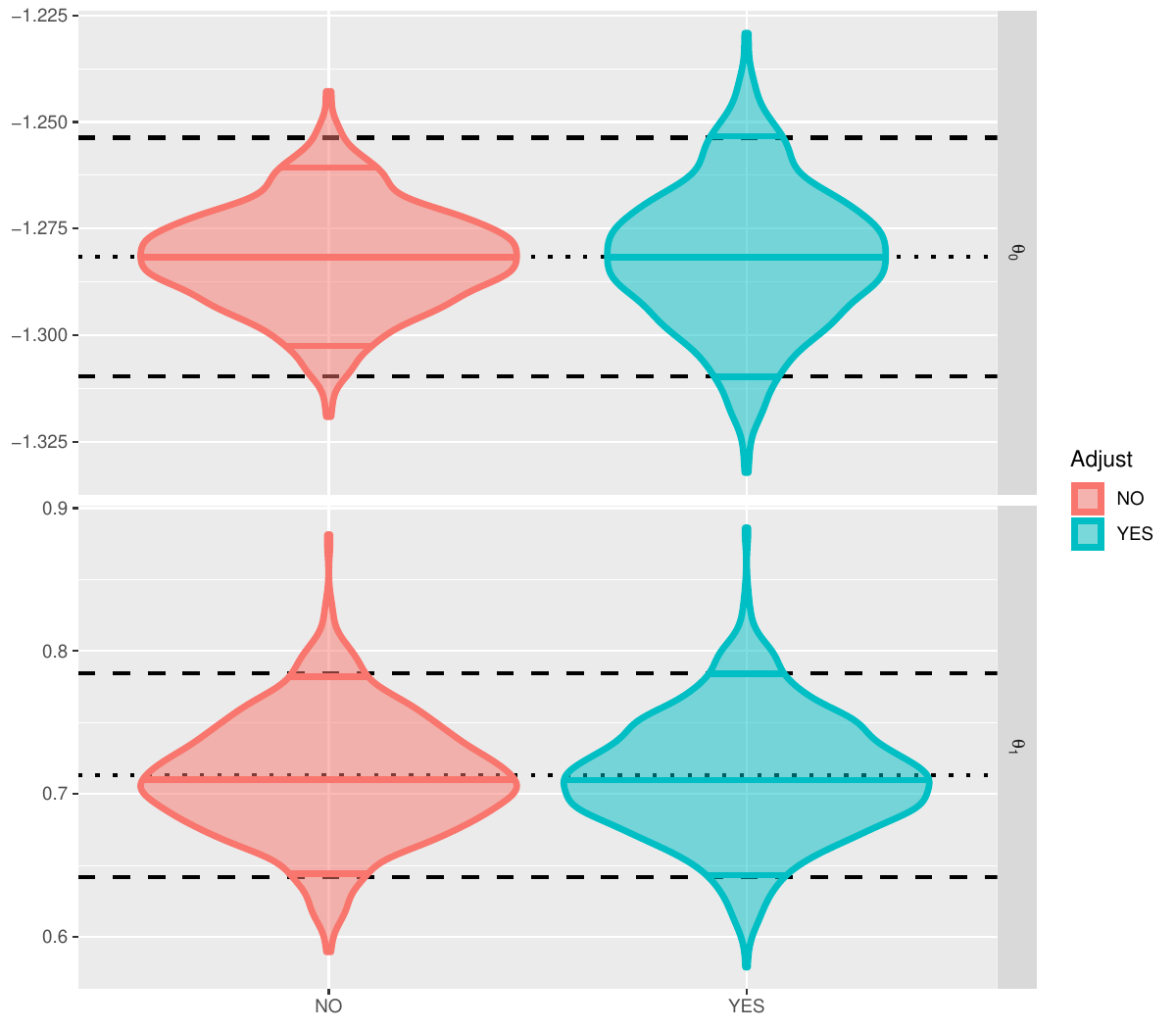}
\caption{Marginal pseudo-posterior sample for the intercept (top) and slope (bottom) for a logistic regression modeling current cigarette smoking by past year major depressive episode based the 2014 National Survey on Drug Use and Health. Unadjusted (left) and adjusted (right) with median and 90\% quantiles (solid bars). Asymptotic normal mean (dotted) and 90\% interval  (dashed) for pseudo-MLE. Created with `ggplot2' \citep{Rggplot2}.}
\label{fig:NSDUHmargin}
\end{figure}

Marginal estimates for $\mbf{\theta}$ agree closely when comparing to the survey-weighted MLE (Figure~\ref{fig:NSDUHmargin}). The covariance structure also matches when comparing the adjusted MCMC samples $\hat{\theta}^{a}_m$ to the pseudo-MLE estimates (Figure~\ref{fig:NSDUHscatter}). Given the large sample size of approximately $42,000$ adults, close agreement is expected. However, we still note potential deviations from asymptotic normality in the pseudo-posterior, which may serve as a tool for model diagnostics that are not available with asymptotic normality methods such as in \cite{Wang2017}.

\FloatBarrier

\section{Discussion}
This work is motivated by the need to apply Bayesian models to survey data. Previous works \citep{2015arXiv150707050S, 2018dep} have demonstrated consistency of the survey-weighted pseudo-posterior for a large class of population models and complex survey designs. However, \citet{2017arXiv171000019N} observe that the resulting posterior intervals can have poor frequentist performance. Insights from the composite likelihood \citep{2009arXiv0911.5357R} and model mis-specification \citep{kleijn2012} literature motivated the development of the theory and adjustment of the asymptotic covariance of the survey-weighted pseudo-posterior. This resulting adjusted pseudo-posterior can then be used for inference in the same manner as the posterior distribution from a simple random sample. It also achieves the same asymptotic properties as the pseudo-likelihood under `design-based' frequentist inference methods. While the results match up well with asymptotic normal methods based on the survey-weighted MLE \citep{Wang2017}, the adjusted pseudo-posterior provides more information with respect to small sample properties. These results allow for modelers to better incorporate informative sample design features into their own analysis models while allowing survey statisticians to incorporate more complex modeling approaches into their analysis and production of official statistics, for example quantile regression and penalized splines \citep{2017pair} and multivariate latent variable models for count data \citep{2015arXiv150707050S}.

Adjustment \ref{eq:adjustment} implemented via Algorithm \ref{algo:R2R1} provides a simple, computationally inexpensive, and effective approach to quantifying and adjusting for the warping of the pseudo-posterior due to unequal weighting and complex sampling dependence between sampling units. Our resampling algorithm eliminates the need to analytically integrate $\mbox{Var}_{P_{\theta_{0}}}\dot{\ell}_{\theta_{0}}$ and thus can be applied to the composite pseudo-likelihood as a more flexible alternative to the modified MCMC approaches presented in \citet{2009arXiv0911.5357R}. Its implementation (Section \ref{app:code}) is straight-forward by leveraging existing software for Bayesian estimation, algorithmic differentiation, and variance estimation via survey replicates.
We note that adjustment \ref{eq:adjustment} is a projection, but does not force the pseudo-posterior variance to equal that of the pseudo-MLE exactly for small-to-moderate samples. Instead, it provides an asymptotic adjustment which allows the analyst to base inference on the sample distribution of the posterior (adjusted by the design effect) rather than using the asymptotic MLE covariance. If the latter is desired, benchmarking to force the posterior samples to exactly match specified mean and covariance can be performed in closed form using a constrained linear projection \citep{ghosh92, Datta2011} or via an iterative Newton-Raphson approach for other constrained projections \citep{bench2013}. This benchmarked pseudo-posterior would exactly match the mean and covariance of the samples from \cite{Wang2017} but would still preserve some small sample properties by not forcing the sampling distribution to be normal.

\bibliography{refs_march2019}
\bibliographystyle{chicago}

\appendix

\section{Example Code}\label{app:code}
We present a working R \citep{R} implementation and the code for the NSDUH example in Section \ref{sec:NSDUH}. The function `cs\_sampling' is a wrapper which takes a Stan model \citep{stan:2015}, computes MCMC draws from the (pseudo) posterior, extracts the gradient function via Rstan \citep{Rstan}, creates a replicate design and estimates the variance of the gradient via the `survey' package \citep{Rsurvey}. We then compute and apply the sandwich adjustment in equation \ref{eq:adjustment}. The resampling method to estimate $J^{\pi}_{\theta}$ in Algorithm \ref{algo:R2R1} corresponds to a special case of the `mrbbootstrap' replication option \citep{Preston09}. For the weighted logistic Stan model, see Appendix B of supplementary information of \cite{2018dep}:  \url{doi:10.1214/18-BA1143SUPP}

\subsection{cs\_sampling}
{\scriptsize
\begin{Verbatim}[frame = single, label = {cs\_sampling.r}, labelposition = all, baselinestretch=1]
#cs_sampling###
require(rstan)
require(survey)
require(plyr)

#take in stan mod, list for stan data, name of parameters sampled,
#survey design object (or rep)
#return sampling object with parameters overwritten

cs_sampling <- function(svydes, mod_stan, par_stan, data_stan,
		ctrl_stan = list(chains = 1, iter = 2000, warmup = 1000, thin = 1),
		rep_design = FALSE, ctrl_rep = list(replicates = 100, type = "mrbbootstrap")){

#run STAN model
print("stan fitting")
out_stan  <- sampling(object = mod_stan, data = data_stan,
                          pars = par_stan,
                          chains = ctrl_stan$chains,
                          iter = ctrl_stan$iter, warmup = ctrl_stan$warmup, thin = ctrl_stan$thin
                          )

#Get posterior mean (across all chains)
par_samps <- extract(out_stan, pars = par_stan, permuted = FALSE)
par_hat <- colMeans(par_samps, dim = 2)#dim = 1 by chain, dim = 2 across chains

#Estimate Hessian
Hhat  <- -1*optimHess(par_hat, gr = function(x){grad_log_prob(out_stan, x)})

#create svrepdesign
if(rep_design == TRUE){svyrep <- svydes
	}else{
	svyrep <- as.svrepdesign(design = svydes, type = ctrl_rep$type, replicates = ctrl_rep$replicates)
}

#Estimate Jhat = Var(gradient)
print("gradient evaluation")
rep_tmp <- withReplicates(design = svyrep, theta = grad_par, stanmod = mod_stan,
						standata = data_stan, par_stan = par_stan, par_hat = par_hat)
Jhat <- vcov(rep_tmp)

#compute adjustment
Hi <- solve(Hhat)
V1 <- Hi%*%Jhat%*%Hi
R1 <- chol(V1)
R2i <- chol(Hi)
R2 <- solve(R2i)
R2R1 <- R2%*%R1

#adjust samples
par_adj <- aaply(par_samps, 1, DEadj, par_hat = par_hat, R2R1 = R2R1, .drop = FALSE)
#matches par_samps if needed

return(list(stan_fit = out_stan, sampled_parms = par_samps, adjusted_parms =par_adj))

}#end of cs_sampling


##helper functions###

##grad_par helper function to nest within withReplicates()
#Stan will pass warnings from calling 0 chains, but still create out_stan object with
#grad_log_prob() method
grad_par <- function(pwts, svydata, stanmod, standata,par_stan,par_hat){
#ignore svydata argument it allows access to svy object data
standata$weights <- pwts
out_stan  <- sampling(object = stanmod, data = standata,
                      pars = par_stan,
                      chains = 0, warmup = 0,
                      )

gradpar <- grad_log_prob(out_stan,par_hat)
return(gradpar)
}#end of grad theta

#helper function to apply matrix rotation
DEadj <- function(par, par_hat, R2R1){
par_adj <- (par - par_hat)%*%R2R1 + par_hat
return(par_adj)
}

\end{Verbatim}
}

\subsection{NSDUH example}
{\scriptsize
\begin{Verbatim}[frame = single, label = {NSDUH\_ISR\_example.r}, labelposition = all, baselinestretch=1]
#National Survey on Drug Use and Health example
#Estimate logistic regression model via Stan
#adjust parameter distribution for complex sample design

setwd("...")
source("cs_sampling.r")#requires rstan

rstan_options(auto_write = TRUE)
mod  <- stan_model('wt_logistic.stan')# compile stan code (doi:10.1214/18-BA1143SUPP)

#import NSDUH data
#https://www.datafiles.samhsa.gov/study-dataset/
	national-survey-drug-use-and-health-2014-nsduh-2014-ds0001-nid16876)#
load(file = "NSDUH_2014.RData")

#subset and clean up the large file#
library(stringr)
#change names to all upper case
names(PUF2014_090718) <- str_to_upper(names(PUF2014_090718))

#QUESTID2: individual ID
#CIGMON: past month smoking
#AMDEY2_U: past year depression (adults)
#CATAG6: Age groups
#ANALWT_C: analysis weights
#VESTR: Variance estimation strata
#VEREP: Variance estimation PSU (nested within strata)

dat14 <- PUF2014_090718[,c("QUESTID2","CIGMON", "AMDEY2_U", "CATAG6", "ANALWT_C", "VESTR", "VEREP")]
rm(PUF2014_090718);gc(); #clean up memory

#subset to adults
dat14 <- dat14[as.numeric(dat14$CATAG6) > 1,]

#normalize weights to sum to sample size
dat14$WTS <- dat14$ANALWT_C*(length(dat14$ANALWT_C)/sum(dat14$ANALWT_C))

#create survey design object#
svy14 <- svydesign(ids = ~VEREP, strata = ~VESTR, weights = ~WTS, data = dat14, nest = TRUE)

#create list of inputs#
X <- model.matrix( ~ AMDEY2_U, data = dat14)
y <- dat14$CIGMON
k   <- dim(X)[2]
n   <- length(y)
weights <- dat14$WTS

#list of inputs
data_stan <- list(y = array(y, dim = n), X = X, k = k, n = n, weights = array(weights, dim = n) )
par_stan <- c("theta") #subset of parameters interested in

#run Stan and adjustment code
set.seed(12345) #set seed to fix output for comparison
mod1 <- cs_sampling(svydes = svy14, mod_stan = mod, par_stan = par_stan, data_stan = data_stan)

##compare to svyglm
svyglm1 <- svyglm(CIGMON ~ AMDEY2_U, design = svy14, family = quasibinomial())

#asymptotic normal 90% ellipse for MLE
library(car)
MLell <- data.frame(ellipse(center = coef(svyglm1),
				shape = vcov(svyglm1),
				radius = sqrt(qchisq(0.90,2)), draw = FALSE),
				Adjust = NA)
names(MLell) <- c("th0", "th1", "Adjust")

#asymptotic normal 90% intervals for MLE
qML0 <- qnorm(c(0.05, 0.95), coef(svyglm1)[1], sqrt(diag(vcov(svyglm1)))[1])
qML1 <- qnorm(c(0.05, 0.95), coef(svyglm1)[2], sqrt(diag(vcov(svyglm1)))[2])

##plot before/after adjustment
library(ggplot2)
#Joint distribution#

datpl <- data.frame(rbind(mod1$sampled_parms[,1,], mod1$adjusted_parms[,1,])
				, as.factor(c(rep("NO", 1000), rep("YES", 1000))))
names(datpl) <- c("th0", "th1", "Adjust")

plt1 <- ggplot(datpl, aes(th0, th1, color = Adjust, shape = Adjust )) +
	geom_point()+
	stat_ellipse(level = 0.90, type = "norm", size = 2)+
	geom_path(data = MLell, aes(th0,th1), size = 2, linetype = "dashed") +
	labs( x= expression(theta [0]), y = expression(theta [1]))
print(plt1)

#marginal distribution
df1 <- data.frame(datpl[,-2], par = 0, ML = coef(svyglm1)[1], q5ML = qML0[1], q95ML = qML0[2])
df2 <- data.frame(datpl[,-1], par = 1, ML = coef(svyglm1)[2], q5ML = qML1[1], q95ML = qML1[2])
names(df1) <- names(df2) <- c("Est", "Adjust", "Par", "ML", "q5ML", "q95ML")

datpl2 <- rbind(df1,df2)

plt2 <- ggplot(datpl2, aes(y= Est, x= Adjust, color = Adjust, fill = Adjust)) +
	geom_hline(aes(yintercept = ML), size = 1, linetype = "dotted")+
	geom_hline(aes(yintercept = q5ML), size = 1, linetype = "dashed")+
	geom_hline(aes(yintercept = q95ML), size = 1, linetype = "dashed")+
	geom_violin(trim=TRUE,draw_quantiles = c(0.05, 0.5, 0.95),alpha=0.5, size = 1.5)+
	facet_grid(Par~., scales = "free", labeller = label_bquote(theta [.(Par)]))+
	labs( x = NULL, y = NULL)
print(plt2)
\end{Verbatim}
}

\section{Proofs}\label{app:proofs}

\subsection{Proof of Theorem~\ref{explimit}  (Asymptotic Normality of the Pseudo-MLE)}
The proof strategy closely follows Theorem 5.23 of \cite{vdv00} where we update the centered and scaled empirical process, $\mathbb{G}_{N_{\nu}}$ to its sampling-weighted extension, $\mathbb{G}^{\pi}_{N_{\nu}}$.  For every random sequence, $h_{N_{\nu}}$, we extend \citet{vdv00} Lemma 19.31 to achieve,
\begin{equation}\label{empexp}
\mathbb{G}^{\pi}_{N_{\nu}}\left(\sqrt{N_{\nu}}\left(\ell_{\theta_{0}+\frac{h_{N_{\nu}}}{\sqrt{N_{\nu}}}} - \ell_{\theta_{0}}\right) - h_{N_{\nu}}^{T}\dot{\ell}_{\theta_{0}}\right) \ipnu 0.
\end{equation}
Conditions ~\nameref{continuity} and ~\nameref{bartlettfirst}, along with
\begin{align}
\mathbb{E}_{P_{\nu}}\left[\mathbb{P}^{\pi}_{N_{\nu}}\ell_{\theta}\right] &= \mathbb{E}_{P_{\nu}}\left[\frac{1}{N_{\nu}}\mathop{\sum}_{i=1}^{N_{\nu}}\frac{\delta_{\nu i}}{\pi_{\nu i}}\ell_{\theta}(\mbf{X}_{i})\right]\\
&= \mathbb{P}_{N_{\nu}}\ell_{\theta}
\end{align}
produces a $0$ mean for the random sequence of Equation~\ref{empexp} with respect to the joint distribution, $(P_{\theta},P_{\nu})$.  By the boundedness requirement for sequence $(\pi_{\nu i}^{-1})$ in Condition~\nameref{bounded}, the Lipschitz condition in Condition~\nameref{continuity} and the dominated convergence theorem, their variance converges to $0$ and the result in Equation~\ref{empexp} is achieved.

Conditions ~\nameref{continuity}, ~\nameref{consistency},~\nameref{bounded} and Corollary 5.53 of \cite{vdv00}, the sequence $h_{N_{\nu}} = \sqrt{N_{\nu}}\left(\theta-\theta_{0}\right)$ is bounded in probability.

We may re-write Equation~\ref{empexp} as,
\begin{equation*}
  N_{\nu}\mathbb{P}^{\pi}_{N_{\nu}}\log\frac{p_{\theta_{0}+\frac{h_{N_{\nu}}}{\sqrt{N_{\nu}}}}}{p_{\theta_{0}}} - h_{N_{\nu}}^{T}\mathbb{G}^{\pi}_{N_{\nu}}\dot{\ell}_{\theta_{0}} - N_{\nu}\mathbb{E}_{P_{\theta_{0}}}\log\frac{p_{\theta_{0}+\frac{h_{N_{\nu}}}{\sqrt{N_{\nu}}}}}{p_{\theta_{0}}} = \smallO_{P_{\theta_{0}},P_{\nu}}(1)
\end{equation*}

From Condition~\nameref{expansion}, we have,
\begin{equation*}
  \mathbb{E}_{P_{\theta_{0}}}\log\frac{p_{\theta_{0}+\frac{h_{N_{\nu}}}{\sqrt{N_{\nu}}}}}{p_{\theta_{0}}} -\frac{1}{2N_{\nu}}h_{N_{\nu}}^{T}H_{\theta_{0}}h_{N_{\nu}} = \smallO_{P_{\theta_{0}}}(1)
\end{equation*}

Substituting this expression above yields,
\begin{equation}\label{LAN}
  N_{\nu}\mathbb{P}^{\pi}_{N_{\nu}}\log\frac{p_{\theta_{0}+\frac{h_{N_{\nu}}}{\sqrt{N_{\nu}}}}}{p_{\theta_{0}}} =
  \frac{1}{2}h_{N_{\nu}}^{T}H_{\theta_{0}}h_{N_{\nu}} + h_{N_{\nu}}^{T}\mathbb{G}^{\pi}_{N_{\nu}}\dot{\ell}_{\theta_{0}}
  + \smallO_{P_{\theta_{0}},P_{\nu}}(1),
\end{equation}
that we recognize as the local asymptotic normality condition of \citet{kleijn2012} (which we will later use to derive the form for the asymptotic covariance matrix of the pseudo-posterior distribution).  Equation~\ref{LAN} is true for both $\hat{h}^{\pi}_{N_{\nu}}$ and $\tilde{h}^{\pi}_{N_{\nu}} = - H_{\theta_{0}}^{-1}\mathbb{G}^{\pi}_{N_{\nu}}\dot{\ell}_{\theta_{0}}$ by Condition~\nameref{consistency}.  The remainder of the proof exactly follows \citet{vdv00} where we separately plug in each of $\hat{h}^{\pi}_{N_{\nu}}$ and $\tilde{h}^{\pi}_{N_{\nu}}$ for $h_{N_{\nu}}$ into Equation~\ref{LAN} to achieve two equivalent equations (up to $\smallO_{P_{\theta_{0},P_{\nu}}}(1)$).  We take the difference between the two equations and complete the square, which produces the result of the theorem.

\subsection{Proof of Theorem~\ref{pseudo-mle} (Aysmptotic Variance of the Pseudo-MLE) }
We begin by constructively expanding the variance with respect to the joint distribution, $(P_{\theta_{0}},P_{\nu})$,
\begin{equation}\label{goalexpr}
\mbox{Var}_{P_{\theta_{0}},P_{\nu}}\{ -H_{\theta_{0}}^{-1}\sqrt{N_{\nu}}\mathbb{P}^{\pi}_{N_{\nu}}\dot{\ell}_{\theta_{0}} \}=
N_{\nu}H_{\theta_{0}}^{-1}\mbox{Var}_{P_{\theta_{0}},P_{\nu}}\mathbb{P}^{\pi}_{N_{\nu}}\dot{\ell}_{\theta_{0}}H_{\theta_{0}}^{-1}.
\end{equation}
We proceed to apply the total variance decomposition to the variance of the random sequence in the middle of the above expression,
\begin{equation}\label{totvar}
  \mbox{Var}_{P_{\theta_{0}},P_{\nu}}\mathbb{P}^{\pi}_{N_{\nu}}\dot{\ell}_{\theta_{0}} = \mbox{Var}_{P_{\theta_{0}}}\mathbb{E}_{P_{\nu}}\left[\mathbb{P}^{\pi}_{N_{\nu}}\dot{\ell}_{\theta_{0}}\mid\mathcal{A}_{\nu}\right]
  + \mathbb{E}_{P_{\theta_{0}}}\mbox{Var}_{P_{\nu}}\left[\mathbb{P}^{\pi}_{N_{\nu}}\dot{\ell}_{\theta_{0}}
  \mid\mathcal{A}_{\nu}\right],
\end{equation}
where, fixing $\nu$, $\mathcal{A}_{\nu}$ denotes the sigma field of information in the population, $U_{\nu}$.  Next, we constructively evaluate each of the two terms.
\begin{subequations}\label{firstterm}
\begin{align}
  \mbox{Var}_{P_{\theta_{0}}}\mathbb{E}_{P_{\nu}}\left[\mathbb{P}^{\pi}_{N_{\nu}}\dot{\ell}_{\theta_{0}}\mid\mathcal{A}_{\nu}\right] & = \mbox{Var}_{P_{\theta_{0}}}\left\{\frac{1}{N_{\nu}}\mathop{\sum}_{i=1}^{N_{\nu}}\frac{\mathbb{E}_{\nu}\left[\delta_{\nu i}\mid\mathcal{A}_{\nu}\right]}{\pi_{\nu i}}\dot{\ell}_{\theta_{0}}(\mbf{X}_{i})\right\} \\
  & = \mbox{Var}_{P_{\theta_{0}}}\left\{\frac{1}{N_{\nu}}\mathop{\sum}_{i=1}^{N_{\nu}}\dot{\ell}_{\theta_{0}}(\mbf{X}_{i})\right\}\\
  & = \frac{1}{N_{\nu}^{2}}\mathbb{E}_{P_{\theta_{0}}}\left\{\mathop{\sum}_{i=1}^{N_{\nu}}\dot{\ell}_{\theta_{0}}(\mbf{X}_{i})\right\}^{2}\\
  & = \frac{1}{N_{\nu}^{2}}\left[\mathop{\sum}_{i=1}^{N_{\nu}} \mathbb{E}_{P_{\theta_{0}}}\dot{\ell}_{\theta_{0}}(\mbf{X}_{i})\dot{\ell}_{\theta_{0}}(\mbf{X}_{i})^{T} + \mathop{\sum}_{i\neq j\in U_{\nu}}\mathbb{E}_{P_{\theta_{0}}}\dot{\ell}_{\theta_{0}}(\mbf{X}_{i})\dot{\ell}_{\theta_{0}}(\mbf{X}_{j})^{T}\right]\\
  &=\frac{1}{N_{\nu}^{2}}\mathop{\sum}_{i=1}^{N_{\nu}} \mathbb{E}_{P_{\theta_{0}}}\dot{\ell}_{\theta_{0}}(\mbf{X}_{i})\dot{\ell}_{\theta_{0}}(\mbf{X}_{i})^{T},
\end{align}
\end{subequations}
where the second term in the second equation from the bottom results because $\mbf{X}_{i}\perp \mbf{X}_{j}$ under $P_{\theta_{0}}$ and by Condition~\nameref{bartlettfirst}.

\begin{subequations}\label{secondterm}
\begin{align}
  \mathbb{E}_{P_{\theta_{0}}}\mbox{Var}_{P_{\nu}}\left[\mathbb{P}^{\pi}_{N_{\nu}}\dot{\ell}_{\theta_{0}}
  \mid\mathcal{A}_{\nu}\right] & =  \mathbb{E}_{P_{\theta_{0}}}\left\{\frac{1}{N_{\nu}^{2}}\mbox{Var}_{P_{\nu}}\left[\mathop{\sum}_{i=1}^{N_{\nu}}\frac{\delta_{\nu i}}{\pi_{\nu i}}\dot{\ell}_{\theta_{0}}\right]\mid\mathcal{A}_{\nu}\right\}\\
  \begin{split}
  & =\frac{1}{N_{\nu}^{2}}\mathbb{E}_{P_{\theta_{0}}}\left\{\mathop{\sum}_{i=1}^{N_{\nu}}\mbox{Var}_{P_{\nu}}\left[\frac{\delta_{\nu i}\mid\mathcal{A}_{\nu}}{\pi_{\nu i}}\right]\dot{\ell}_{\theta_{0}}(\mbf{X}_{i})\dot{\ell}_{\theta_{0}}(\mbf{X}_{i})^{T} \right.\\
  & \left.+ \mathop{\sum}_{i\neq j\in U_{\nu}}\mbox{Cov}_{P_{\nu}}\left[\frac{\delta_{\nu i}\delta_{\nu j}\mid\mathcal{A}_{\nu}}{\pi_{\nu i}\pi_{\nu j}}\right]\dot{\ell}_{\theta_{0}}(\mbf{X}_{i})\dot{\ell}_{\theta_{0}}(\mbf{X}_{j})^{T}\right\}
  \end{split}\\
  \begin{split}\label{covterms2}
  &=\frac{1}{N_{\nu}^{2}}\mathop{\sum}_{i=1}^{N_{\nu}}\mathbb{E}_{P_{\theta_{0}}}\left\{\left[\frac{1}{\pi_{\nu i}}-1\right]\dot{\ell}_{\theta_{0}}(\mbf{X}_{i})\dot{\ell}_{\theta_{0}}(\mbf{X}_{i})^{T}\right\}\\
  & + \frac{1}{N_{\nu}^{2}}\mathop{\sum}_{i\neq j\in U_{\nu}}\mathbb{E}_{P_{\theta_{0}}}\left\{\left[\frac{\pi_{\nu ij}}{\pi_{\nu i}\pi_{\nu j}}-1\right]\dot{\ell}_{\theta_{0}}(\mbf{X}_{i})\dot{\ell}_{\theta_{0}}(\mbf{X}_{j})^{T}\right\}
  \end{split}\\
  \begin{split}\label{boundedcov}
  &\leq\frac{1}{N_{\nu}^{2}}\mathop{\sum}_{i=1}^{N_{\nu}}\mathbb{E}_{P_{\theta_{0}}}\left\{\left[\frac{1}{\pi_{\nu i}}-1\right]\dot{\ell}_{\theta_{0}}(\mbf{X}_{i})\dot{\ell}_{\theta_{0}}(\mbf{X}_{i})^{T}\right\}\\
  & + \max\{1,\gamma-1\}\frac{1}{N_{\nu}^{2}}\mathop{\sum}_{i\neq j\in U_{\nu }}\biggl\vert\mathbb{E}_{P_{\theta_{0}}}\left\{\dot{\ell}_{\theta_{0}}(\mbf{X}_{i})\dot{\ell}_{\theta_{0}}(\mbf{X}_{j})^{T}\right\}\biggr\vert
  \end{split}\\
  &=\frac{1}{N_{\nu}^{2}}\mathop{\sum}_{i=1}^{N_{\nu}}\mathbb{E}_{P_{\theta_{0}}}\left\{\left[\frac{1}{\pi_{\nu i}}-1\right]\dot{\ell}_{\theta_{0}}(\mbf{X}_{i})\dot{\ell}_{\theta_{0}}(\mbf{X}_{i})^{T}\right\}.\label{finalcov}
\end{align}
\end{subequations}
The sequence, $-1 \leq \displaystyle\left[\frac{\pi_{\nu ij}}{\pi_{\nu i}\pi_{\nu j}}-1\right]$, in Equation~\ref{covterms2} is bounded from above by $\displaystyle\left[\frac{1}{\pi_{\nu i}}-1\right] \leq (\gamma - 1)$ by Condition~\nameref{bounded}. See \citet{2018dep} for more details.  The second expression in Equation~\ref{boundedcov} exactly equals $0$ by the independence of $\mbf{X}_{i}$ and $\mbf{X}_{j}$ ($\forall~ i \neq j \in U_{\nu}$) under $P_{\theta_{0}}$ and by Condition~\nameref{bartlettfirst}. Since the second expression in Equation~\ref{boundedcov} is bounded from above by $0$, it exactly equals $0$ (for all $\nu\in\mathbb{Z}^{+},~i\neq j\in U_{\nu}$), producing the equality in Equation~\ref{finalcov}.  Equation~\ref{covterms2} results from the following computations:
\begin{align*}
  \mbox{Var}_{P_{\nu}}\left[\frac{\delta_{\nu i}\mid\mathcal{A}_{\nu}}{\pi_{\nu i}}\right] & = \mathbb{E}_{P_{\nu}}\left[\frac{\delta_{\nu i}\mid\mathcal{A}_{\nu}}{\pi_{\nu i}}\right]^{2}-\left[\mathbb{E}_{P_{\nu}}\frac{\delta_{\nu i}\mid\mathcal{A}_{\nu}}{\pi_{\nu i}}\right]^{2} \\
   & = \frac{1}{\pi_{\nu i}} - 1\\
  \mbox{Cov}_{P_{\nu}}\left[\frac{\delta_{\nu i}\delta_{\nu j}\mid\mathcal{A}_{\nu}}{\pi_{\nu i}\pi_{\nu j}}\right] & = \mathbb{E}_{P_{\nu}}\left[\frac{\delta_{\nu i}\delta_{\nu j}\mid\mathcal{A}_{\nu}}{\pi_{\nu i}\pi_{\nu j}}\right]-\mathbb{E}_{P_{\nu}}\left[\frac{\delta_{\nu i}\mid\mathcal{A}_{\nu}}{\pi_{\nu i}}\right]\mathbb{E}_{P_{\nu}}\left[\frac{\delta_{\nu j}\mid\mathcal{A}_{\nu}}{\pi_{\nu j}}\right] \\
  & = \frac{\pi_{\nu ij}}{\pi_{\nu i}\pi_{\nu j}} - 1
\end{align*}

We plug in the results for Equations~\ref{firstterm} and \ref{secondterm} back into Equation~\ref{totvar},
\begin{subequations}\label{varnugget}
\begin{align}
\begin{split}
  N_{\nu}\mbox{Var}_{P_{\theta_{0}},P_{\nu}}\mathbb{P}^{\pi}_{N_{\nu}}\dot{\ell}_{\theta_{0}} & =\frac{1}{N_{\nu}}\mathop{\sum}_{i=1}^{N_{\nu}} \mathbb{E}_{P_{\theta_{0}}}\dot{\ell}_{\theta_{0}}(\mbf{X}_{i})\dot{\ell}_{\theta_{0}}(\mbf{X}_{i})^{T} \\
  & + \frac{1}{N_{\nu}}\mathop{\sum}_{i=1}^{N_{\nu}}\mathbb{E}_{P_{\theta_{0}}}\left\{\left[\frac{1}{\pi_{\nu i}}-1\right]\dot{\ell}_{\theta_{0}}(\mbf{X}_{i})\dot{\ell}_{\theta_{0}}(\mbf{X}_{i})^{T}\right\}
  \end{split}\\
  & \leq \gamma J_{\theta_{0}}
\end{align}
\end{subequations}
and the result is achieved.

\subsection{Proof of Theorem~\ref{pseudo-posterior} (Asymptotic Distribution of the Pseudo-Posterior) }
The proof strategy is the same as \citet{kleijn2012} where they first prove the assertion on any two sets of random sequences, $(h_{N_{\nu}},g_{N_{\nu}}) \in K$, where $K\in\mathbb{R}^{d}$ is an arbitrary compact set.  They then extend the result to a sequence of balls, $K_{N_{\nu}}$, centered on $0$ with increasing radii, $M_{N_{\nu}}\uparrow\infty$.  We extend their strategy by updating notation to incorporate the $(\delta_{\nu i},\pi_{\nu i})$, where $\pi_{\nu i} = \mbox{Pr}\{\delta_{\nu i} = 1\}$, governed by the sampling design distribution, $P_{\nu}$, such that our result applies for $(P_{\theta_{0}},P_{\nu})$, jointly.  Recall that we have the local asymptotic normality result,
\begin{equation}\label{compactLAN}
  N_{\nu}\mathbb{P}^{\pi}_{N_{\nu}}\log\frac{p_{\theta_{0}+\frac{h_{N_{\nu}}}{\sqrt{N_{\nu}}}}}{p_{\theta_{0}}} =
  \frac{1}{2}h_{N_{\nu}}^{T}H_{\theta_{0}}h_{N_{\nu}} + h_{N_{\nu}}^{T}\mathbb{G}^{\pi}_{N_{\nu}}\dot{\ell}_{\theta_{0}}
  + \smallO_{P_{\theta_{0}},P_{\nu}}(1),
\end{equation}
from the proof of Theorem~\ref{explimit} under Conditions~\nameref{continuity}, ~\nameref{expansion}, ~\nameref{bartlettfirst} and ~\nameref{bounded}.  Define the sample-weighted empirical loglikelihood ratio,
\begin{equation}\label{pseudoratio}
  s_{N_{\nu}}^{\pi}(h) = N_{\nu}\mathbb{P}_{N_{\nu}}^{\pi}\log\frac{P_{\theta_{0}}+\frac{h}{\sqrt{N}_{\nu}}}{P_{\theta_{0}}},
\end{equation}
and let $\DD^{\pi}_{N_{\nu},\theta_{0}} = H_{\theta_{0}}^{-1}\mathbb{G}^{\pi}_{N_{\nu}}\dot{\ell}_{\theta_{0}}$.  Plugging into Equation~\ref{compactLAN}, we achieve,
\begin{equation*}
  s_{N_{\nu}}^{\pi}(h_{N_{\nu}}) = h_{N_{\nu}}^{T}H_{\theta_{0}}\DD^{\pi}_{N_{\nu},\theta_{0}} - \frac{1}{2}h_{N_{\nu}}^{T}H_{\theta_{0}}h_{N_{\nu}} + \smallO_{P_{\theta_{0}},P_{\nu}}(1).
\end{equation*}
Let $\phi_{N_{\nu}}$ denote the normal distribution, $\mathcal{N}\left(\DD^{\pi}_{N_{\nu},\theta_{0}},H_{\theta_{0}}^{-1}\right)$ and define the sequence of random functions,
\begin{equation}\label{posteriorratio}
  f^{\pi}_{N_{\nu}}\left(g_{N_{\nu}},h_{N_{\nu}}\right) = \left( 1 - \frac{\phi_{N_{\nu}}(h_{N_{\nu}})s^{\pi}_{N_{\nu}}(g_{N_{\nu}})\pi_{N_{\nu}}(g_{N_{\nu}})}{\phi_{N_{\nu}}(g_{N_{\nu}})s^{\pi}_{N_{\nu}}(h_{N_{\nu}})\pi_{N_{\nu}}(h_{N_{\nu}})}\right)_{+}.
\end{equation}
Plugging into the logarithm of Equation~\ref{posteriorratio} for $s^{\pi}_{N_{\nu}}(\cdot)$ and $\phi_{N_{\nu}}(\cdot)$, where for any $(h_{N_{\nu}},g_{N_{\nu}}) \in K$, the prior ratio, $\pi_{N_{\nu}}(g_{N_{\nu}})/\pi_{N_{\nu}}(h_{N_{\nu}}) \rightarrow 1$ as $\nu\uparrow\infty$, we achieve:
\begin{subequations}\label{ratio}
\begin{align}
  &\log \left(\frac{\phi_{N_{\nu}}(h_{N_{\nu}})s^{\pi}_{N_{\nu}}(g_{N_{\nu}})\pi_{N_{\nu}}(g_{N_{\nu}})}{\phi_{N_{\nu}}(g_{N_{\nu}})s^{\pi}_{N_{\nu}}(h_{N_{\nu}})\pi_{N_{\nu}}(h_{N_{\nu}})}\right) = \\
  \begin{split}
   & = \left(g_{N_{\nu}}-h_{N_{\nu}}\right)^{T}H_{\theta_{0}}\DD^{\pi}_{N_{\nu},\theta_{0}} + \frac{1}{2}h_{N_{\nu}}^{T}H_{\theta_{0}}h_{N_{\nu}} -  \frac{1}{2}g_{N_{\nu}}^{T}H_{\theta_{0}}g_{N_{\nu}} + \smallO_{P_{\theta_{0}},P_{\nu}}(1)\\
   & -\frac{1}{2}\left(h_{N_{\nu}}-\DD^{\pi}_{N_{\nu},\theta_{0}}\right)^{T}H_{\theta_{0}}\left(h_{N_{\nu}}-\DD^{\pi}_{N_{\nu},\theta_{0}}\right) + \frac{1}{2}\left(g_{N_{\nu}}-\DD^{\pi}_{N_{\nu},\theta_{0}}\right)^{T}H_{\theta_{0}}\left(g_{N_{\nu}}-\DD^{\pi}_{N_{\nu},\theta_{0}}\right) \end{split}\\
   & = \smallO_{P_{\theta_{0}},P_{\nu}}(1),
\end{align}
\end{subequations}
as $\nu\uparrow\infty$.  Conclude that
\begin{equation}\label{fgoeszero}
  \mathop{sup}_{g,h\in K} f^{\pi}_{N_{\nu}}(g,h) \mathop{\rightarrow}^{P_{\theta_{0}},P_{\nu}} 0,
\end{equation}
as $\nu\uparrow\infty$.  Define $\Xi_{N_{\nu}}$ as the event that $\Pi^{\pi}_{N_{\nu}}(K) > 0$.
Define \[\Pi^{\pi,K}_{N_{\nu}}\left(B\mid \mbf{X}_{\nu},\bm{\delta}_{\nu}\right) = \Pi^{\pi}_{N_{\nu}}\left(h\in B\mid \mbf{X}_{\nu},\bm{\delta}_{\nu}\right) / \Pi^{\pi}_{N_{\nu}}\left(K\mid \mbf{X}_{\nu},\bm{\delta}_{\nu}\right)\]
to the posterior mass truncated to the compact space, $K$, and similarly for $\Phi^{K}_{N_{\nu}}$.   Fix (any) $\eta > 0$ and define the sequence of events, $\Omega_{N_{\nu}} = \left\{\mathop{\sup}_{g,h\in K} f^{\pi}_{N_{\nu}}(g,h) \leq \eta\right\}$. Construct the inequality,
\begin{equation}\label{boundonK}
  \mathbb{E}_{P_{\theta_{0}},P_{\nu}}\norm{\Pi^{\pi,K}_{N_{\nu}} - \Phi^{K}_{N_{\nu}}}\mathbf{1}_{\Xi_{N_{\nu}}} \leq \mathbb{E}_{P_{\theta_{0}},P_{\nu}}\norm{\Pi^{\pi,K}_{N_{\nu}} - \Phi^{K}_{N_{\nu}}}\mathbf{1}_{\Omega_{N_{\nu}} \cap \Xi_{N_{\nu}}} + 2\mathbb{E}_{P_{\theta_{0}},P_{\nu}}\left( \Xi_{N_{\nu}} \backslash \Omega_{N_{\nu}} \right),
\end{equation}
where the total variation normal, $\norm{\cdot}$, is bounded above by $2$ and the second on the right-hand side is $\smallO\left(1\right)$ from Equation~\ref{fgoeszero}. Since $\norm{P-Q} = 2\int (1-p/q)^{+}dQ$, we may expand the first term on the right-hand side,
\begin{align}
&\frac{1}{2}\mathbb{E}_{P_{\theta_{0}},P_{\nu}}\norm{\Pi^{\pi,K}_{N_{\nu}} - \Phi^{K}_{N_{\nu}}}\mathbf{1}_{\Omega_{N_{\nu}} \cap \Xi_{N_{\nu}}} \\
&\leq \mathbb{E}_{P_{\theta_{0}},P_{\nu}} \int \left( 1 - \frac{\phi_{N_{\nu}}(h)s^{\pi}_{N_{\nu}}(g)\pi_{N_{\nu}}(g)}{\phi_{N_{\nu}}(g)s^{\pi}_{N_{\nu}}(h)\pi_{N_{\nu}}(h)}\right)_{+}
d\Phi^{K}_{N_{\nu}}(g)d\Pi^{\pi,K}_{N_{\nu}}(h)\mathbf{1}_{\Omega_{N_{\nu}} \cap \Xi_{N_{\nu}}}\\
&\leq \mathbb{E}_{P_{\theta_{0}},P_{\nu}}\int \mathop{\sup}_{g,h\in K} f^{\pi}_{N_{\nu}}(g,h)\mathbf{1}_{\Omega_{N_{\nu}} \cap \Xi_{N_{\nu}}}d\Phi^{K}_{N_{\nu}}(g)d\Pi^{\pi,K}_{N_{\nu}}(h) \leq \eta.
\end{align}

The proof next follows \citet{kleijn2012} to expand the result on a compact $K$ to compact sets, $\left(K_{N_{\nu}}\right)_{\nu}$ of balls with radii $M_{N_{\nu}}\uparrow\infty$, which provides the result for $\mathbb{R}^{d}$ in the limit of $\nu$.  From Theorem~\ref{explimit}, we have:
\begin{equation}
\hat{h}_{\pi,N_{\nu}} = \sqrt{N}_{\nu}\left(\hat{\theta}_{\pi,N_{\nu}}-\theta_{0}\right) = -\DD^{\pi}_{N_{\nu},\theta_{0}} + \smallO_{P_{\theta_{0}},P_{\nu}}(1),
\end{equation}
and the stated result is achieved with a rescaling and shift since the total variation norm is invariant to rescalings and shifts.

\end{document}